\documentclass[pre,twocolumn,usletter,showpacs,superscriptaddress]{revtex4}

\usepackage{amsmath}
\usepackage{amsfonts}
\usepackage{amssymb}

\usepackage [ansinew] {inputenc}
\usepackage{graphicx}
\usepackage{natbib}

\begin{document}
\title{Nonlocal and global dynamics of cellular automata: \\ A theoretical computer arithmetic for real continuous maps}

\author{Vladimir Garc\'{\i}a-Morales}

\affiliation{Institute for Advanced Study - Technische Universit\"{a}t M\"{u}nchen, Lichtenbergstr. 2a, D-85748 Garching, Germany}
%\affiliation{Nonequilibrium Chemical Physics - Physics Department - Technische Universit\"{a}t M\"{u}nchen, James-Franck-Str. 1, D-85748 Garching, Germany}

\begin{abstract}
\noindent A digit function is presented which provides the $i$th-digit in base $p$ of any real number $x$.  By means of this function, formulated within $\mathcal{B}$-calculus, the local, nonlocal and global dynamical behaviors of cellular automata (CAs) are systematically explored and universal maps are derived for the three levels of description. None of the maps contain any freely adjustable parameter and they are valid for any number of symbols in the alphabet $p$ and neighborhood range $\rho$. A discrete general method to approximate any real continuous map in the unit interval by a CA on the rational numbers $\mathbb{Q}$ (Diophantine approximation) is presented. This result leads to establish a correspondence between the qualitative behavior found in bifurcation diagrams of real nonlinear maps and the Wolfram classes of CAs. The method is applied to the logistic map, for which a logistic CA is derived. The period doubling cascade into chaos is interpreted as a sequence of global cellular automata of Wolfram's class 2 leading to Class 3 aperiodic behavior. Class 4 behavior is also found close to the period-3 orbits.
 \end{abstract}
\maketitle

\section{Introduction}

One of the challenges of chaos theory is to construct maps or classes of maps that correctly represent a given dynamical system \cite{McCauley}. Von Neumann \cite{Neumann} observed that any discretization of a system of differential equations for computation is a replacement of the system by an automaton, and he suggested addressing the statistics generated by ``artificial automata'' to gain insight into the workings of real or model systems that are too complex for ordinary analysis (he specifically thought in automata that can reproduce themselves). Von Neumann's artificial automata, later called cellular automata (CAs) \cite{Neumann, Codd, Wolfram1, Wolfram2, PhysicaD, Adamatzky, Ilachinski, VGM1, VGM2, VGM3, group, Ban, Blanchard, Kari, Mcintosh, Wuensche, Chua1, Israeli1, Israeli2} provide an important pathway to understand how local and global dynamics are related in systems of increasingly larger complexity. The fact that these systems evolve on a discrete spacetime and that they can only be on a finite number of dynamical states make them ideal, minimalistic objects to study the complexity that purely arises from local or nonlocal interactions of range $\rho$ as well as the global behavior associated with them.

CA can serve as a basis for a fully discrete method \cite{Wolfram1}, a ``theoretical computer arithmetic'' \cite{McCauley, MC2, MCPAL1, MCPAL2}, which leads to a more general formulation of deterministic chaos in terms of symbolic dynamics \cite{Lind}. Indeed, the effect of a positive Lyapunov exponent in the chaotic regime is quite elegantly captured in terms of symbolic dynamics, since it shows how a flow of information takes place on finite strings of symbols from the less significant digits (where the truncation is made) to the most significant ones. Although irrational numbers \cite{Niven} are taken for granted in theoretical physics and are important in the study of nonlinear dynamical systems as well (KAM tori and Siegel discs provide good examples) finite precision is a fact that cannot be avoided neither in computation nor in experiment \cite{McCauley, Wolfram1} and the question whether a fully discrete dynamical system can approximate to arbitrary precision a real map or a system of differential equations is an important one. Together with the methods of symbolic dynamics \cite{Lind}, fixed-point arithmetics employing strings of digits from a finite alphabet to a sufficiently high precision leads not only to clarify the essentials of chaotic dynamics, but also to provide more accurate quantitative results for the chaotic pseudo-orbits than does the usual floating-point arithmetic \cite{McCauley}.  Even when the shadowing lemma (Bowen-Anosov lemma) \cite{McCauley, Ott} can be invoked as a reason why one needs not to take care to compute correctly a trajectory when the system is chaotic, if one is to understand the results of computation \cite{Wolfram1} the latter lemma is of no help. Symbolic dynamics can then be put to work to solve that task. Furthermore, floating-point arithmetic can introduce numerical errors that may misleadingly turn a periodic orbit into a seemingly aperiodic one \cite{McCauley}. 

In this article we advocate McCauley's approach to nonlinear dynamical systems \cite{McCauley} and we provide a general mathematical framework which accomplishes the ``theoretical computer arithmetic'' that he advanced through some examples (focusing on Baker maps and Bernoulli shifts). Since fixed-point arithmetic and symbolic dynamics are consistently implemented by means of CA (see last chapter in \cite{McCauley}), we pursue here the elucidation of the systematic general means to change from a real map to the description provided by CA. This is achieved through a series of results for the non-local and global dynamics of CA that we present here with the hope (in view of the main result) that they might constitute tools not only for the study of CA but also for the analysis of dynamical systems in general. The approach is based on $\mathcal{B}$-calculus \cite{VGM1, VGM2, VGM3} and on a function introduced in Lemma 1 which works as a kind of CA-transform, allowing any number to be replaced by a string of digits in any integer base to a certain precision. We first derive equivalent forms for the universal map implementing the local dynamics of CA. Then, we establish which CA rules in computational space act as shift operators. These play then a prominent role in the subsequent discussion, where we then derive the universal maps for the nonlocal and the global dynamics for CA. The three levels of description are thus provided by three interrelated universal CA maps: the local map (description at the level of site values), the non-local map (at the level of entire neighborhoods of sites or strings of contiguous neighborhoods) and the global map (the dynamics at the level of the entire system) also called global characteristic function. Once the connection between the three different descriptions is established, a CA is formulated to approximate with arbitrary precision any nonlinear map on the real unit interval.

The outline of this paper is as follows. In Sec. \ref{local} we briefly review $\mathcal{B}$-calculus and our previous results for the local dynamics of CA \cite{VGM1} and then introduce an auxiliary function (Lemma 1) which allows any number in base 10 to be converted to base $p$. We obtain as well some new results for the local dynamics of CA that are useful for the following sections, obtaining CA codes for those rules in computational space that implement the shift-operators. In Sec. \ref{nonlocal} we derive the universal map for the nonlocal dynamics which governs the evolution of the neighborhood values and we prove in a corollary that the neighborhood dynamics exhibited by the nonlocal map takes place on a de Bruijn graph. In Sec. \ref{global} we derive the universal characteristic function for the global dynamics of CA. Some tools, useful to characterize global CA behavior, are derived as well from this expression, and are then discussed and illustrated with examples. Global CA, whose neighborhood range equals the total system size and for which, therefore, the nonlocal and the global dynamics coincide, are then introduced. It is then shown that the set of global shift operators form an abelian group under composition. Finally, the main result on the approximation to arbitrary accuracy of a real map by a global CA is established (Theorem 6) and illustrated with the logistic map. This example shows that there is a one-to-one correspondence between Wolfram CA classes of complexity and the qualitative behavior exhibited by nonlinear maps on the real unit interval. Some conclusions are then presented summarizing the main results.

\section{Local dynamics of cellular automata (CA)}
\label{local}

\begin{figure*}
\includegraphics[width=0.5\textwidth, angle=270]{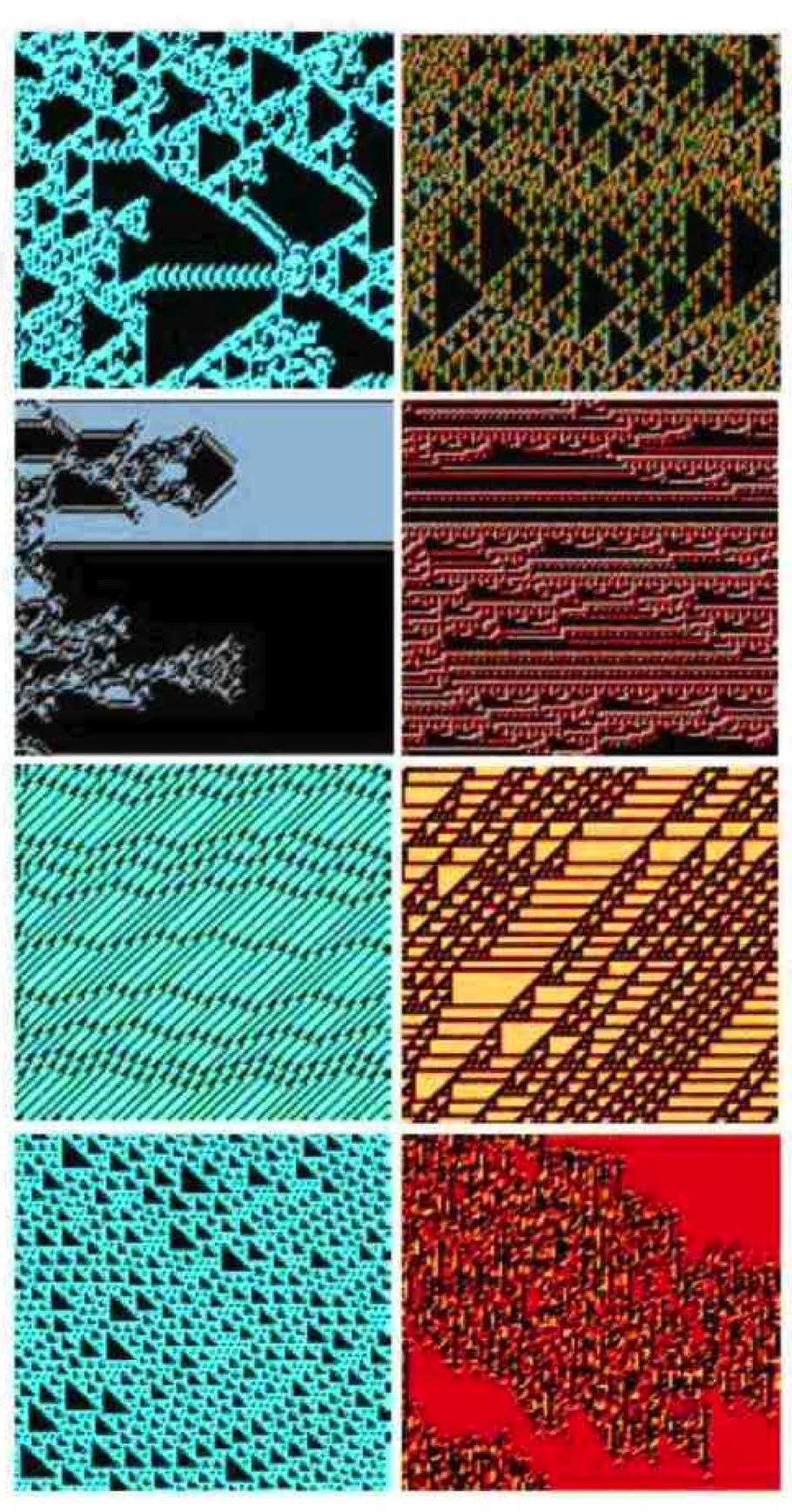}
\caption{(Color online) Spatiotemporal evolution of some 1D CA rules obtained from Eq. (\ref{CA}) for an arbitrary initial condition. From top to bottom and right to left the following rules are shown: $^1 110 _2 ^1$, $^2 29649 _2 ^1$, $^2 52T _2 ^2$, $^3 88T _2 ^3$, $^3 9334T _3 ^4$, $^2 51649 _2 ^1$, $^0 8322 _3 ^1$, $^1 93340T_5 ^1$. (In rules with totalistic code, Eq. (\ref{CAT}) is used instead). In each panel, time flows from top to bottom and space spans along the horizontal direction. Shown is a window $100\times 100$ in each case. } \label{cas}  
\end{figure*}

Let us consider a 1D ring containing a total number of $N_{s}$ sites. An input is given as initial condition in the form of a vector $\mathbf{x}_{0}=(x_{0}^{1},...,x_{0}^{N_{s}})$. Each of the $x^{i}_{0}$ is an integer in the range $0$ through $p-1$ where superindex $i \in [1, N_{s}]$ specifies the position of the site on the 1D ring. At each $t$ the vector $\mathbf{x}_{t}=(x_{t}^{1},...,x_{t}^{N_{s}})$ specifies the state of the CA.  Periodic boundary conditions are considered so that $x_{t}^{N_{s}+1}=x_{t}^{1}$ and $x_{t}^{0}=x_{t}^{N_{s}}$. Let $x_{t+1}^{i}$ be taken to denote the value of site $i$ at time step $t+1$.  Formally, its dependence on the values at the previous time step is given through the mapping $x_{t+1}^{i}=\ ^{l}R_{p}^{r}(x_{t}^{i+l},\ ... x_{t}^{i}, \ ...,x_{t}^{i-r} )$, which we abbreviate as $x_{t+1}^{i}=\ ^{l}R_{p}^{r}(x_{t}^{i})$ with the understanding that the function on the r.h.s depends on all site values within the neighborhood, with range $\rho=l+r+1$, which contains the site $i$ updated at the next time ($l$ and $r$ denote the number of cells to the left and to the right of site $i$ respectively). We take the convention that $i$ increases to the left. The integer number $n$ in base 10, which runs between $0$ and $p^{r+l+1}-1$, indexes all possible neighborhood values coming from the different configurations of site values. Each of these configurations compares to the dynamical configuration reached by site $i$ and its $r$ and $l$ first-neighbors at time $t$ and given by 
\begin{equation}
n_{t}^{i}=\sum_{k=-r}^{l}p^{k+r}x_{t}^{i+k} \label{NV}
\end{equation}
We will refer to this latter quantity often as the \emph{neighborhood value}. The possible outputs $a_{n}$ for each configuration $n$ are also integers $\in[0,p-1]$. An integer number $R$ can then be given in base 10 to fully specify the rule $^{l}R_{p}^{r} $ as 
\begin{equation}
R \equiv \sum_{n=0}^{p^{r+l+1}-1}a_{n}p^{n}. \label{RWolf}
\end{equation}
With all these specifications we have the following universal map \cite{VGM1}
\begin{equation}
x_{t+1}^{i}=\ ^{l}R_{p}^{r}(x_{t}^{i})=\sum_{n=0}^{p^{r+l+1}-1}a_{n}\mathcal{B}\left(n-\sum_{k=-r}^{l}p^{k+r}x_{t}^{i+k}, \frac{1}{2} \right) \label{CA}
\end{equation}
where $\mathcal{B}(x,\epsilon)$ is the boxcar function, 
\begin{equation}
\mathcal{B}(x,\epsilon)=\frac{1}{2}\left(\frac{x+\epsilon}{|x+\epsilon|}-\frac{x-\epsilon}{|x-\epsilon|}\right) \label{d10}
\end{equation} 
which returns 1 when $|x| < \epsilon$ and 0 otherwise. This function is the building block of $\mathcal{B}$-calculus \cite{VGM1}. In this paper we shall always take $\epsilon=1/2$ and we shall abbreviate $\mathcal{B}(x,\epsilon)$ simply by $\mathcal{B}(x)$, i.e.
\begin{equation}
\mathcal{B}(x)\equiv \frac{1}{2}\left(\frac{x+\frac{1}{2}}{|x+\frac{1}{2}|}-\frac{x-\frac{1}{2}}{|x-\frac{1}{2}|}\right) \label{d1}
\end{equation} 
Eq. (\ref{CA}) describes the local dynamics of \emph{all first-order-in-time deterministic CA rules in 1D} with no freely adjustable parameters: the $p^{r+l+1}$ coefficients $a_{n}$ directly specify the dynamical rule. For example, for Wolfram's rule $^{1}110_{2}^{1}$,  $\mathbf{a}=(a_{0},a_{1},...a_{7})=(0,1,1,1,0,1,1,0)$ (see Fig. 2 in \cite{VGM1}, where all above notation is clarified).

%Although $\mathcal{B}$-calculus \cite{VGM1} makes the specification of the number $R$ no longer necessary (since all information of the CA rule is already contained in the vector $\mathbf{a}=(a_0, a_1,...a_{p^{r+l+1}-1})$ and in the parameters $p$, $l$ and $r$) it is retained here to simplify the discussion regarding most elementary CA rules.

CA whose output value depend on the sum over the previous neighborhood values (and not from any other specific detail of the site-values configuration) are called totalistic.These CA rules constitute a subset of all CA rules described by Eq. (\ref{CA}). For totalistic CA the following simpler map \cite{VGM1} can be used
\begin{equation}
x_{t+1}^{i}=\sum_{s=0}^{\rho(p-1)}a_{s}\mathcal{B}\left(s-\sum_{k=-r}^{l}x_{t}^{i+k}\right) \label{CAT}
\end{equation}
where $\rho=l+r+1$ and, again, each $a_{s}$ is an integer between $0$ and $p-1$ like the inputs and the output of the rule, which is now labelled as $^{l}RT_{p}^{r}$, with $R=\sum_{s=0}^{\rho(p-1)}a_{s}p^{s}$.  In Fig. \ref{cas} the spatiotemporal evolution of some CAs obtained from Eqs. (\ref{CA}) and (\ref{CAT}) are shown, giving just a visual impression of the wide variety of different complex behaviors that can arise during CA evolution. 

Within $\mathcal{B}$-calculus, modular arithmetics can be formulated \cite{VGM2}. The quotient of the division of the non-negative integer number $m$ by the natural number $p$ is given by
\begin{equation}
\left \lfloor\frac{m}{p}\right \rfloor=\sum_{j=0}^{m}\sum_{k=0}^{p-1}j\mathcal{B}\left(m-jp-k\right) \label{quot}
\end{equation}
and the remainder by
\begin{equation}
m-p\left \lfloor\frac{m}{p}\right \rfloor=\sum_{j=0}^{m}\sum_{k=0}^{p-1}k\mathcal{B}\left(m-jp-k\right)= m \mod p\label{rem}
\end{equation}
where the brackets $\left \lfloor ... \right \rfloor$ denote the lower nearest integer (floor) function. 
The r. h. s. of Eqs. (\ref{quot}) and Eq. (\ref{rem}) scan all relevant integers $j$ and $k$ to find the pair $(j,k)$ that satisfies $m=jp+k$. 

%We thus also have
%\begin{equation}
%\lceil\frac{m}{p}\rceil=\sum_{j=0}^{m}\sum_{k=0}^{p-1}(j+1)\mathcal{B}\left(m-jp-k\right) \label{quotp1}
%\end{equation}
%with $\lceil ... \rceil$ denoting the upper nearest integer (ceiling) function.
%Since we often will need to convert numbers from the decimal base to base $p$ we describe now a simplest mathematical expression to achieve this specific goal.

A \emph{representation in radix $p>1, p \in \mathbb{N}$ of a real number $x \in \mathbb{R}$}  has the form
\begin{equation}
x=a_{N}p^{N-1}+a_{N-1}p^{N-2}+\ldots+a_{1}+a_{0}p^{-1}+a_{-1}p^{-2}+\ldots \label{bexpa}
\end{equation}
where the $a_{i} \in \mathbb{Z}$ ($i \in \mathbb{Z}$) satisfy $0 \le a_{i} \le p-1$. We have 
\begin{equation}
\left \lfloor \frac{x}{p^{i-1}} \right \rfloor=a_{N}p^{N-i}+\ldots +a_{i+1}p+a_{i}
\end{equation} 
whence, by subtracting 
\begin{equation}
p \left \lfloor \frac{x}{p^{i}} \right \rfloor=a_{N}p^{N-i}+\ldots +a_{i+1}p 
\end{equation}
we obtain 
\begin{equation}
a_{i}=\left \lfloor \frac{x}{p^{i-1}} \right \rfloor-p\left \lfloor \frac{x}{p^{i}} \right \rfloor \equiv \mathbf{d}_{p}(i,x) \label{cucuA}
\end{equation}
This is  the \emph{digit function} $\mathbf{d}_{p}(i,x)$: It gives the $i$-th digit of $x$ in radix $p$ \cite{VGM1}. This function plays a central role in a new recent formulation of quantum mechanics based on the principle of least radix economy \cite{QUANTUMPAPER}.

We now prove some important properties in the case $x=A$ being a non-negative integer (all them can be easily extended to $x$ real as well). 
%
%
%Let $A$ denote a non-negative integer. If we write this number in base $p$ we have
%\begin{eqnarray}
%A=\sum_{i=1}^{\left \lfloor \log_{p}A \right \rfloor+1} a^{(i)} p^{i-1} \label{p2ten}
%\end{eqnarray}
%where $a^{(i)}$ is the $i$ digit of the number in base $p$. We now define these $a^{(i)}$ as the integer function $\mathbf{d}_p(i,A)$ ($\equiv a^{(i)}$). 
%~\\

\noindent \emph{Lemma 1: If $x=A$ is a non-negative integer, the digit function $\mathbf{d}_{p}(i,A)$ satisfies:}
\begin{eqnarray}
&& \mathbf{d}_p(i,A) =\sum_{j=0}^{\left \lfloor A/p^{i-1}\right \rfloor}\sum_{k=0}^{p-1}k\mathcal{B}\left(\left \lfloor \frac{ A}{p^{i-1}} \right \rfloor  -jp-k\right) \nonumber \\ && \label{boxdig}\\
&& \mathbf{d}_p(i,A) = 0 \qquad \forall i > \left \lfloor \log_{p}A \right \rfloor+1 \label{bound} \\
&& \sum_{i=1}^{\left \lfloor \log_{p}A \right \rfloor+1} p^{i-1} \mathbf{d}_p(i,A) = A \label{iden} \\
&& \lim_{p \to \infty} \mathbf{d}_p(i,A) = A\ \mathcal{B}\left(i-1\right)  \label{limi} \\
&& \mathbf{d}_p(i,p^{m-1}) = \mathbf{d}_p(m,p^{i-1})= \mathcal{B}\left(i-m\right)  \label{krone} \\
&& \mathbf{d}_p(i,A) = \sum_{j=1}^{B}\mathbf{d}_p(i,j)\mathbf{d}_p(j,p^{A-1}) \quad (\forall B \ge A) \label{cuttie}
\end{eqnarray}

\noindent \emph{Proof:} For $i=1$ Eq. (\ref{cucuA}) takes the form
\begin{eqnarray}
\mathbf{d}_p(1,A) &=& A-p\left \lfloor \frac{ A}{p} \right \rfloor   \label{eq1}
\end{eqnarray}
which corresponds to the remainder upon dividing $A$ by $p$ [see also Eq. (\ref{rem})]. When $\mathbf{d}_p(1,A)=0$ we say that the number $A$ is divisible by $p$. The quotient is given by $\left \lfloor A/p \right \rfloor$, Eq. (\ref{quot}). After $i-1$ divisions, the quotient is $\left \lfloor A/p^{i-1} \right \rfloor$, because of the nesting property of the floor function \cite{Graham}. A further division by $p$ yields the quotient  $\left \lfloor A/p^{i} \right \rfloor$ and the remainder of that division is just Eq. (\ref{cucuA}) which, can be rewritten, by using Eq. (\ref{rem}) to be put in the form of Eq. (\ref{boxdig}).

To prove Eq. (\ref{bound}) note that since $\left \lfloor \log_{p}A \right \rfloor \le  \log_{p}A  \le \left \lfloor \log_{p}A \right \rfloor+1$ we have $p^{\left \lfloor \log_{p}A \right \rfloor} \le  A  \le p^{\left \lfloor \log_{p}A \right \rfloor+1}$ and therefore $\left \lfloor A/p^{i} \right \rfloor=0 \ \forall i \ge \left \lfloor \log_{p}A \right \rfloor+1$. From the definition Eq. (\ref{cucuA}) this in turn implies Eq. (\ref{bound}). 

The proof of the useful Eq. (\ref{iden}) proceeds by noting that  
\begin{eqnarray}
&&\sum_{i=1}^{\left \lfloor \log_{p}A \right \rfloor+1} p^{i-1} \mathbf{d}_p(i,A) \nonumber \\
&&= \sum_{i=1}^{\left \lfloor \log_{p}A \right \rfloor+1} p^{i-1} \left(\left \lfloor \frac{A}{p^{i-1}} \right \rfloor -p\left \lfloor \frac{A}{p^{i}} \right \rfloor \right) \nonumber \\
&&= \sum_{i=1}^{\left \lfloor \log_{p}A \right \rfloor+1}  \left(p^{i-1}\left \lfloor \frac{A}{p^{i-1}} \right \rfloor -p^{i}\left \lfloor \frac{A}{p^{i}} \right \rfloor \right) \nonumber \\
&&= p^{0}\left \lfloor \frac{A}{p^{0}} \right \rfloor-p^{\left \lfloor \log_{p}A \right \rfloor+1}\left \lfloor \frac{A}{p^{\left \lfloor \log_{p}A \right \rfloor+1}} \right \rfloor \nonumber \\
&&=A
\end{eqnarray} 
where we have used that $\left \lfloor A/p^{\left \lfloor\log_{p}A \right \rfloor+1} \right \rfloor=0$, a result obtained in proving Eq. (\ref{bound}) (see above). Eqs. (\ref{limi}) and (\ref{krone}) are direct consequences of Eq. (\ref{iden}). Eq. (\ref{cuttie}) follows from Eqs. (\ref{bound}) and (\ref{krone}).
$\Box$

Eq. (\ref{cucuA}) is the mathematical equivalent of the algorithm to find the base $p$ representation of a number in base $10$ \cite{Knuth}. An example where this formula is already useful is provided by the Wolfram coding of CA rules: the integer $R$ in base 10 is given by Eq. (\ref{RWolf}). Therefore, we, conversely, have $a_n=\mathbf{d}_p(n+1,R)$. 

~\\

\noindent \emph{Theorem 1: For the non-negative integers $x_{t}^{i+k} \in [0,p-1]$, the universal CA map, Eq. (\ref{CA}) can be equivalently written as}
\begin{equation}
x_{t+1}^{i}=\sum_{n=0}^{p^{r+l+1}-1}a_{n} \mathbf{d}_p\left(\sum_{k=-r}^{l} p^{k+r}x_{t}^{i+k},p^{n-1}\right) \label{calto}
\end{equation}
\emph{or as} 
\begin{equation}
x_{t+1}^{i}=\sum_{n=0}^{p^{r+l+1}-1}a_{n}\prod_{k=-r}^{l}\mathcal{B}\left(\mathbf{d}_p(k+r+1,n)-x_{t}^{i+k}\right) \label{calt}
\end{equation}

\noindent \emph{Proof}: Eq. (\ref{calto}) follows directly from Eq. (\ref{CA}) and Eq. (\ref{krone}). 
To prove Eq. (\ref{calt}) note that, since $0 \le n \le p^{l+r+1}$, by using Eq. (\ref{iden}) we have
\begin{eqnarray}
&&n=\sum_{k=1}^{l+r+1} p^{k-1} \mathbf{d}_p(k,n) \nonumber \\
&&=\sum_{k=-r}^{l} p^{k+r} \mathbf{d}_p(k+r+1,n)
\end{eqnarray} 
Thus, from Eq. (\ref{CA}) 
\begin{eqnarray}
x_{t+1}^{i}&&=\sum_{n=0}^{p^{r+l+1}-1}a_{n}\times  \label{CApr} \\
&&\times \mathcal{B}\left(\sum_{k=-r}^{l}p^{k+r}\left[\mathbf{d}_p(k+r+1,n)-x_{t}^{i+k}\right]\right) \nonumber
\end{eqnarray}
The r. h. s. of the latter equation implies that $\forall k$ the $(k+r+1)$th digit of $n$ in base $p$ must also match the corresponding $x_{t}^{i+k}$ i.e.
\begin{eqnarray}
&&\mathcal{B}\left(\sum_{k=-r}^{l}p^{k+r}\left[\mathbf{d}_p(k+r+1,n)-x_{t}^{i+k}\right]\right) \nonumber \\
&&=\prod_{k=-r}^{l}\mathcal{B}\left(\mathbf{d}_p(k+r+1,n)-x_{t}^{i+k}\right)
\end{eqnarray}
By replacing this result in Eq. (\ref{CApr}) we obtain Eq. (\ref{calt}) thus proving the theorem. $\Box$   
~\\

\noindent \emph{Corollary (Wolfram rules): For the 256 Wolfram rules $^{1}R^{1}_{2}$, Eq. (\ref{calt}) reduces to the map}
\begin{eqnarray} %twocolumn, version 2
&&x_{t+1}^{i}=a_{0}(1-x_{t}^{i+1})(1-x_{t}^{i})(1-x_{t}^{i-1})+ a_{1}x_{t}^{i-1}\times\nonumber \\
&&\times (1-x_{t}^{i+1})(1-x_{t}^{i})+a_{2}x_{t}^{i}(1-x_{t}^{i+1})(1-x_{t}^{i-1}) + \nonumber \\
&&+a_{3}x_{t}^{i}x_{t}^{i-1}(1-x_{t}^{i+1})+a_{4}x_{t}^{i+1}(1-x_{t}^{i})(1-x_{t}^{i-1})+ \nonumber \\
&&+a_{5}x_{t}^{i+1}x_{t}^{i-1}(1-x_{t}^{i})+a_{6}x_{t}^{i+1}x_{t}^{i}(1-x_{t}^{i-1})+ \nonumber \\
&&+a_{7}x_{t}^{i+1}x_{t}^{i}x_{t}^{i-1} \\
&&=a_{0}+(a_{1}-a_{0})x_{t}^{i-1}+(a_{2}-a_{0})x_{t}^{i}+(a_{4}-a_{0})x_{t}^{i+1}+  \nonumber \\
&&+ (a_{3}-a_{2}-a_{1}+a_{0})x_{t}^{i}x_{t}^{i-1} + \nonumber \\
&&+ (a_{5}-a_{4}-a_{1}+a_{0})x_{t}^{i+1}x_{t}^{i-1} + \nonumber \\
&&+ (a_{6}-a_{4}-a_{2}+a_{0})x_{t}^{i+1}x_{t}^{i}+ \nonumber \\
&&+ (a_{7}-a_{6}-a_{5}+a_{4}-a_{3}+a_{2}+a_{1}-a_{0})x_{t}^{i+1}x_{t}^{i}x_{t}^{i} \nonumber
 \label{CAW} 
\end{eqnarray}

\noindent \emph{Proof}: Since $p=2$, $l=r=1$, Eq. (\ref{calt}) takes the form
\begin{equation}
x_{t+1}^{i}=\sum_{n=0}^{7}a_{n}\prod_{k=-1}^{1}\mathcal{B}\left(\mathbf{d}_2(k+r+1,n)-x_{t}^{i+k}\right) \label{caltp}
\end{equation}
For $p=2$ all $\mathbf{d}_2(k+r+1,n)$ and $x_{t}^{i+k}$ can only be either 0 or 1. By noting that $\mathcal{B}(1-y)=y$ and $\mathcal{B}(y)=1-y$ for an integer variable $y \in [0,1]$ we have
\begin{eqnarray}
&&\mathcal{B}\left(\mathbf{d}_2(k+r+1,n)-x_{t}^{i+k}\right) \\
&&=x_{t}^{i+k}\mathcal{B}\left(\mathbf{d}_2(k+r+1,n)-1\right)+ \nonumber \\
&&+(1-x_{t}^{i+k})\mathcal{B}\left(\mathbf{d}_2(k+r+1,n)\right) \nonumber
\end{eqnarray}
By replacing this expression in Eq. (\ref{caltp}) and writing explicitly each term in the sum we obtain Eq. (\ref{CAW}). The proof of this corollary constitutes an alternative to the one presented in \cite{VGM1}. $\Box$
~\\
Eq. (\ref{CAW}) constitutes a map for \emph{all} 256 Wolfram rules that are the subject of major expositions \cite{Wolfram1, Chua1}. It contains \emph{no freely adjustable parameters}: the numbers $a_{0}$ to $a_{7}$ are either zeroes or ones and serve to specify the rule. Our map thus contrasts with recent proposals where non-trivial free parameters are needed in order to model each particular Wolfram CA map (see e.g. \cite{Ozelim}, Eq.(11) and pp. 291-294).

\noindent \emph{Theorem 2 (Shift rules): For a CA rule $^{l}R^{r}_{p}$ with $a_{n}=\mathbf{d}_p(m,n)=\left \lfloor \frac{n}{p^{m-1}} \right \rfloor -p\left \lfloor \frac{n}{p^{m}} \right \rfloor$ (where $m$ is an integer $1 \le m \le \ l+r+1$) and, therefore, with code $R=\sum_{n=0}^{p^{\rho}-1}\mathbf{d}_p(m,n)p^{n}$, the universal CA map Eq. (\ref{CA}) takes the simple form}
\begin{equation}
x_{t+1}^{i}=x_{t}^{i+m-r-1} \label{ls}
\end{equation}

\noindent \emph{Proof}: From Eq. (\ref{calto}) we have
\begin{eqnarray}
x_{t+1}^{i}&=&\sum_{n=0}^{p^{r+l+1}-1}a_{n}\mathbf{d}_p\left(\sum_{k=-r}^{l} p^{k+r}x_{t}^{i+k},p^{n-1}\right) \nonumber \\
&=&\sum_{n=0}^{p^{r+l+1}-1}\mathbf{d}_p(m,n)\mathbf{d}_p\left(\sum_{k=-r}^{l} p^{k+r}x_{t}^{i+k},p^{n-1}\right) \nonumber \\
&=&\sum_{n=0}^{p^{r+l+1}-1}\mathbf{d}_p(m,n)\mathbf{d}_p\left(n, p^{\sum_{k=-r}^{l} p^{k+r}x_{t}^{i+k}-1}\right) \nonumber \\
&=&\mathbf{d}_p\left(m,\sum_{k=-r}^{l} p^{k+r}x_{t}^{i+k}\right)= x_{t}^{i+m-r-1} \label{curesu}
\end{eqnarray}
where Eqs.(\ref{krone}) and (\ref{cuttie}), with $B=p^{l+r+1}-1 \ge  \sum_{k=-r}^{l} p^{k+r}x_{t}^{i+k}$, have been used. $\Box$
~\\

\noindent \emph{Corollary (Identity rule): For a CA rule $^{l}R^{r}_{p}$ with $a_{n}=\left \lfloor \frac{n}{p^{r}} \right \rfloor -p\left \lfloor \frac{n}{p^{r+1}} \right \rfloor$, the universal CA map Eq. (\ref{CA}) becomes}
\begin{equation}
x_{t+1}^{i}=x_{t}^{i} \label{id}
\end{equation}

\noindent \emph{Proof}: This results follows directly from Theorem 2, for the specific case $m=r+1$. $\Box$

\begin{figure}
\includegraphics[width=0.26\textwidth, angle=270]{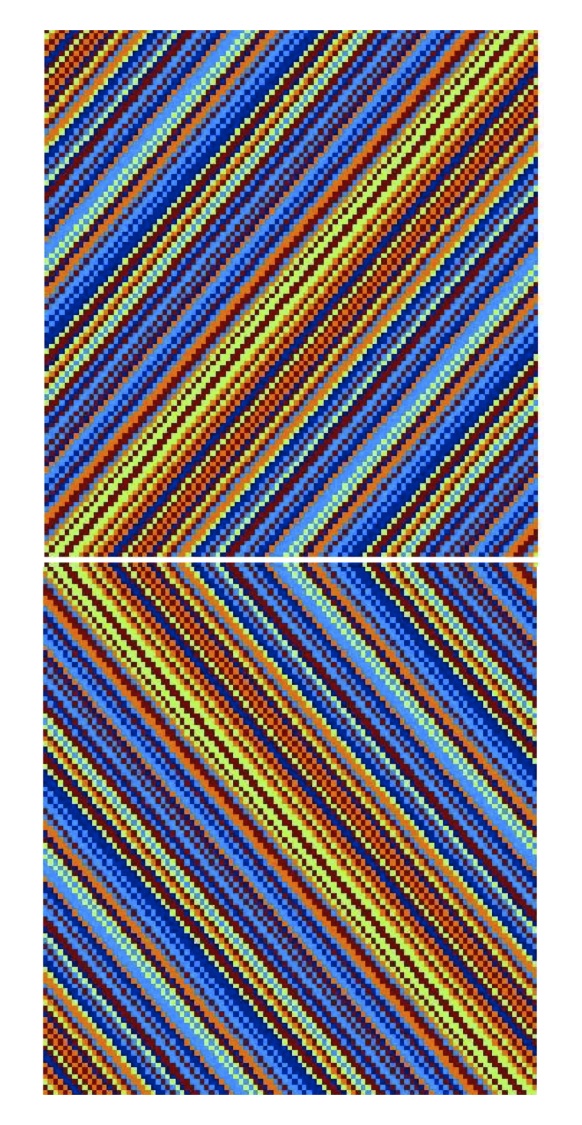}
\caption{(Color online) Spatiotemporal evolution of the shift rules $^{1}R^{1}_{5}$ with codes $\sum_{n=0}^{124}\left(n-5\left \lfloor \frac{n}{5}\right \rfloor\right)5^{n}$ (left) and $\sum_{n=0}^{124}\left \lfloor \frac{n}{25}\right \rfloor5^{n}$ (right) starting from random initial conditions which are the mirror image of each other. Time flows from top to bottom. Shown is a window $[100\times 100]$.} \label{shifts}  
\end{figure}
 
As an example, let us find all shift rules among Wolfram 256 rules $^{1}R^{1}_{2}$. From Theorem 2 we observe that there are three possibilities for $m$ ($m \in [1, 3]$). The resulting rules have vectors $a_{n}=n -2\left \lfloor \frac{n}{2} \right \rfloor$ for ($m=1$), $a_{n}=\left \lfloor \frac{n}{2} \right \rfloor -2\left \lfloor \frac{n}{4} \right \rfloor$ ($m=2$) and $a_{n}=\left \lfloor \frac{n}{4} \right \rfloor -2\left \lfloor \frac{n}{8} \right \rfloor=\left \lfloor \frac{n}{4} \right \rfloor$ for ($m=3$), where $n \in [0,p^{l+r+1}-1]=[0,7]$ in all cases. These rules have thus vectors $\mathbf{a}=(a_{0},a_{1},...a_{7})=(0,1,0,1,0,1,0,1)$, $(0,0,1,1,0,0,1,1)$ and $(0,0,0,0,1,1,1,1)$ which correspond, respectively to rules $^{1}170^{1}_{2}$, $^{1}204^{1}_{2}$ and $^{1}240^{1}_{2}$. By replacing these $a_{n}$ sets in Eq. (\ref{CAW}), we obtain $x_{t+1}^{i}=x_{t}^{i-1}$, $x_{t+1}^{i}=x_{t}^{i}$ and $x_{t+1}^{i}=x_{t}^{i+1}$, respectively, thus confirming again that these rules are the left-shift, identity and right-shift Wolfram rules, respectively. 

%These rules are indeed long well-known to implement the Bernoulli shift (see \cite{Wolfram2, Wolfram3, McCauley}).
 
We illustrate Theorem 2 with a further example, by finding the codes for the shift rules within $^{1}R^{1}_{5}$. Note that there are $5^{5^{3}}=5^{125}$ possible rules with $\rho=3$ and $p=5$. The rules that implement the left-shift, identity and right-shift have, from Theorem 2, codes $R=\sum_{n=0}^{124}\left(n-5\left \lfloor \frac{n}{5}\right \rfloor\right)5^{n}$, $R=\sum_{n=0}^{124}\left(\left \lfloor \frac{n}{5}\right \rfloor-5\left \lfloor \frac{n}{25}\right \rfloor\right)5^{n}$ and $R=\sum_{n=0}^{124}\left \lfloor \frac{n}{25}\right \rfloor5^{n}$ respectively. In Fig. \ref{shifts} the spatiotemporal evolution of  the left-shift  (Fig. \ref{shifts} left) and the right-shift (Fig. \ref{shifts} right) rules is shown (starting from a random initial condition that is the same in both cases after reflection). The dynamical behavior of each of these rules mirrors the one of the other, showing not only that they implement the left and right shifts but also that both rules belong to the same equivalence class under reflection \cite{VGM2}. Because of their significance for all what follows, we introduce a new notation for these rules, to single them out in CA space, and define them as operators.
~\\

\noindent \emph{Definition 1 (Shift operator): We define the shift operator as}
\begin{equation}
\overbrace{^{l}T_{p}^{r}}^{m} \equiv \ ^{l}R^{r}_{p}
\end{equation}
\emph{where the CA rule $^{l}R^{r}_{p}$ has rule vector with components given by Theorem 2 as} 
 \begin{equation}
 a_{n}=\mathbf{d}_p(m,n)=\left \lfloor \frac{n}{p^{m-1}} \right \rfloor -p\left \lfloor \frac{n}{p^{m}} \right \rfloor
 \end{equation}
\emph{($1 \le m \le l+r+1$). We thus have, from Theorem 2}
\begin{equation}
\overbrace{^{l}T_{p}^{r}}^{m}(x_{t}^{i})=x_{t}^{i+m-r-1} \label{oper}
\end{equation}

\noindent \emph{Example:} The Wolfram rules implementing the shift to the left, the identity and the shift to the right are respectively denoted by: 

\noindent $\overbrace{^{1}T_{2}^{1}}^{1}$ (rule $^{1}170^{1}_{2}$), $\overbrace{^{1}T_{2}^{1}}^{2}$ (rule $^{1}204^{1}_{2}$) and $\overbrace{^{1}T_{2}^{1}}^{3}$ (rule $^{1}240^{1}_{2}$) 

%\begin{eqnarray}
%^{1}170^{1}_{2}(x_{t}^{i})& \equiv &\overbrace{^{1}T_{2}^{1}}^{1}(x_{t}^{i}) = x_{t}^{i-1} \\
%^{1}204^{1}_{2}(x_{t}^{i})& \equiv &\overbrace{^{1}T_{2}^{1}}^{2}(x_{t}^{i}) = x_{t}^{i} \\
%^{1}240^{1}_{2}(x_{t}^{i})& \equiv &\overbrace{^{1}T_{2}^{1}}^{3}(x_{t}^{i}) = x_{t}^{i+1} \label{theop}
%\end{eqnarray}
 
~\\ 
The following lemma can be verified by using the above definition.
~\\

 \noindent \emph{Lemma 2: The shift operator satisfies the following recurrence}
\begin{equation}
\overbrace{^{l}T_{p}^{r}}^{m}(x_{t}^{i})=\overbrace{^{l}T_{p}^{r}}^{r+2}(\overbrace{^{l}T_{p}^{r}}^{m-1}(x_{t}^{i})) \label{recurri}
\end{equation} 

Since global translation invariance on the ring holds \cite{VGM2}, the following property also follows. 
~\\

 \noindent \emph{Lemma 3: The shift operator commutes with any CA rule $^{l}R^{r}_{p}$, i.e. one has}
\begin{equation}
\overbrace{^{l}T_{p}^{r}}^{m}[^{l}R^{r}_{p}(x_{t}^{i})]=\ ^{l}R^{r}_{p}[\overbrace{^{l}T_{p}^{r}}^{m} (x_{t}^{i})]=
x_{t+1}^{i+m-r-1}. \label{commu}
\end{equation} 

This latter Lemma is equivalent to Proposition 1.5.7, p.17 in \cite{Lind} under the isomorphism that exists between a 1D CA on an infinite ring ($N_s \to \infty$) and a sliding block code with memory $r$ and anticipation $l$ (see Definition 1.5.1 p. 15 in \cite{Lind}). That the shift operator commutes with any CA rule is also a well-known major result in CA theory \cite{Hedlund}.

\section{Nonlocal dynamics of CA}
\label{nonlocal}

The results in the previous section, all based on Eq. (\ref{CA}) and the definition in Eq. (\ref{cucuA}), concern the behavior of each site value $x_{t}^{i}$ with time, as a function of the values of the neighboring sites $x_{t}^{i+l}$, $x_{t}^{i+l-1}$, ..., $x_{t}^{i-r+1}$, $x_{t}^{i-r}$. We can now ask how the neighborhood value, given by Eq. (\ref{NV}), evolves in time as a function of the neighboring neighborhoods. It is clear that the consistency of CA evolution on the ring demands certain constraints on the contiguity and overlapping of neighborhoods.  If we think in terms of a graph, we can take each possible neighborhood as a vertex on the graph, the edges connecting overlapping neighborhoods in the direction of $i$ increasing to the left. Such directed graph is well known in the literature as a de Bruijn graph \cite{dB, Good, Golomb} and it has been fruitfully applied to CA in some seminal works \cite{Jen1, Jen2, Martin} (this approach is explained in detail in \cite{Mcintosh}).

In this section, we derive a universal map for the nonlocal (neighborhood) dynamics of CA, proving as well in a corollary that \emph{the neighborhood of any CA dynamics, as implemented by the universal map, takes place on a de Bruijn graph}.
~\\

%\noindent \emph{Definition 2: Let $n^{i}$ and $n^{i+1}$ denote neighborhoods. The matrix $\mathbf{b}$ with components $b_{n^{i}n^{i+1}}$ given by}
%\begin{equation}
%b_{n^{i}n^{i+1}}=\prod_{k=1}^{\left \lfloor \log_{p} n^{i} \right \rfloor}\mathcal{B}\left(\mathbf{d}_{p}(k,n^{i+1})-\mathbf{d}_{p}(k+1,n^{i}), \frac{1}{2}\right) \label{dBm}
%\end{equation}
%\emph{is a de Bruijn matrix. $b_{n^{i}n^{i+1}}$ is equal to unity when both neighborhoods are connected and equal to zero otherwise.}  ~\\

\noindent \emph{Definition 2: Let $n$ and $n'$ denote arbitrary neighborhood values. The matrix $\mathbf{b}$ with components $b_{nn'}$ given by}
\begin{equation}
b_{nn'}=\prod_{k=1}^{\left \lfloor \log_{p} n \right \rfloor}\mathcal{B}\left(\mathbf{d}_{p}(k,n')-\mathbf{d}_{p}(k+1,n), \frac{1}{2}\right) \label{dBm}
\end{equation}
\emph{is a de Bruijn matrix. The de Bruijn graph is the directed graph with all possible neighborhood values as vertices (i.e. it has $p^{l+r+1}$ vertices) and whose forward connections are given by Eq. (\ref{dBm}): $b_{nn'}$ is equal to unity when both neighborhoods are connected with an edge $n \to n'$ and equal to zero otherwise.}  ~\\

\noindent \emph{Lemma 4: Let $n^i=\sum_{k=1}^{l+r+1}p^{k-1}\mathbf{d}_{p}(k,n^i)$ denote the vertex number in base 10 (decimal representation of a neighborhood value around $i$) on a de Bruijn graph. Vertex $n^i$ is  connected (i.e., we have $b_{n^i n^{i+1}}=1$) to all vertices $n^{i+1}$ that satisfy }  
\begin{equation}
n^{i+1}=p^{l+r}\mathbf{d}_{p}(l+r+1,n^{i+1})+\sum_{k=1}^{l+r}p^{k-1}\overbrace{^{l}T_{p}^{r}}^{r+2}(\mathbf{d}_{p}(k,n^{i})) \label{vertidin}
\end{equation}

\noindent \emph{Proof:} We note first that, since $n=n^{i}$ in Eq. (\ref{dBm}), $\left \lfloor \log_{p} n^{i} \right \rfloor =l+r$. We note also that, from Eqs. (\ref{curesu}) and Eq. (\ref{oper})
\begin{equation}
\overbrace{^{l}T_{p}^{r}}^{m}(\mathbf{d}_{p}(k,n))=\mathbf{d}_{p}(k+m-r-1,n)
\end{equation} 
Thus, we have, since $1 \le k \le l+r$ in Eq. (\ref{dBm})
\begin{eqnarray}
\mathbf{d}_{p}(k,n^{i+1}) &=& \overbrace{^{l}T_{p}^{r}}^{r+2}(\mathbf{d}_{p}(k,n^{i})) \\
&=& \mathbf{d}_{p}(k+1,n^{i})
\end{eqnarray}
And thus, because this latter expression is satisfied for all $k$ values $\in [1, \left \lfloor \log_{p} n^{i} \right \rfloor]$, from Eq. (\ref{dBm}) it follows that $b_{n^i n^{i+1}}=1$. In Eq. (\ref{vertidin}) there is a term dependent on the non-negative integer $\mathbf{d}_{p}(l+r+1,n^{i+1}) \in [0,p-1]$ and thus there are $p$ edges going out from vertex $n^i$ to the $p$ different nodes $n^{i+1}$. $\Box$

Lemma 4 shows that knowledge of the shift operators for given $l$, $r$ and $p$ allow to construct the relevant de Bruijn graph with $p^{l+r+1}$ vertices. We have, furthermore, the following result.
~\\

 \noindent \emph{Lemma 5: The following relationship holds}
\begin{equation}
\overbrace{^{l}T_{q}^{r}}^{m}(n_{t}^{i})=\sum_{k=-r}^{l}p^{k+r}\overbrace{^{l}T_{p}^{r}}^{m}(x_{t}^{i+k}) \label{usyful}
\end{equation}
\emph{with $q=p^{l+r+1}$ with $n_{t}^{i}=\sum_{k=-r}^{l}p^{k+r}x_{t}^{i+k}$ for each $i \in [1,N_{s}]$.}

\noindent \emph{Proof:} On one hand, we have
\begin{equation}
\overbrace{^{l}T_{q}^{r}}^{m}(n_{t}^{i})=n_{t}^{i+m-r-1}
\end{equation}
on the other
\begin{equation}
\sum_{k=-r}^{l}p^{k+r}\overbrace{^{l}T_{p}^{r}}^{m}(x_{t}^{i})=\sum_{k=-r}^{l}p^{k+r}x_{t}^{i+k+m-r-1}
\end{equation}
Thus, by redefining index $i \to i-m+r+1$ on both expressions (since global translation invariance holds and the position on the ring is given modulo $N_{s}$) we obtain the desired result. $\Box$

The following theorem is the main result of this section and emphasizes the importance of shift operators in CA space. 
~\\

\noindent \emph{Theorem 3: The neighborhood value $n_t^{i}=\sum_{k=-r}^{l}p^{k+r}x_t^{i}$, obtained from a CA rule $^{l}R^{r}_{p}(x_t^{i})$ with code $R=\sum_{n=0}^{p^{l+r+1}-1}a_n p^n$ satisfies}
\begin{equation}
^{l}R^{r}_{p}(x_t^{i})\ =\ ^{0}R'^{0}_{q}(n_t^{i}) \label{reslem5}
\end{equation}
\emph{with $q = p^{l+r+1}$ and $R'= \sum_{n=0}^{q-1}a_n q^n$. The neighborhood value evolves according to the following map}
\begin{equation}
\boxed{n_{t+1}^{i}=\sum_{k=1}^{l+r+1}p^{k-1}\overbrace{^{l}T_{p}^{r}}^{k}[^{0}R'^{0}_{q}(n_t^{i})]} 
\label{NLC}
\end{equation}
%\emph{and a string of $M$ connected neighborhoods $s_t^{i}=\sum_{k'=1}^{M}p^{k'-1}n_t^{i+k'-1}$ evolves as}
%\begin{equation}
%s_{t+1}^{i}=\sum_{k'=1}^{M}\sum_{k=1}^{l+r+1}p^{k+k'-2}\overbrace{^{l}T_{q}^{r}}^{k+k'-1}[^{0}R'^{0}_{q}(n_t^{i})] 
%\label{reslem7}
%\end{equation}

\noindent \emph{Proof:} Eq. (\ref{reslem5}) follows directly from Eq. (\ref{CA}), since, by using that $q \equiv p^{l+r+1}$ and  $n_t^{i}=\sum_{k=-r}^{l}p^{k+r}x_t^{i}$, we have
\begin{equation}
^{l}R^{r}_{p}(x_t^{i})=x_{t+1}^{i}=\sum_{n=0}^{q-1}a_{n}\mathcal{B}\left(n-n_t^i\right)=\ ^{0}R'^{0}_{q}(n_t^{i}) \label{CAl}
\end{equation}
which is Eq. (\ref{reslem5}). To prove Eq. (\ref{NLC}), we observe that, from Eq. (\ref{CAl}), we have
\begin{eqnarray}
&&n_{t+1}^{i} = \sum_{k=-r}^{l}p^{k+r}x_{t+1}^{i+k} = \sum_{k=-r}^{l}p^{k+r}\ ^{0}R'^{0}_{q}(n_t^{i+k}) \nonumber \\
&&= \sum_{k=1}^{l+r+1}p^{k-1}\ ^{0}R'^{0}_{q}(n_t^{i+k-r-1}) \nonumber \\ &&= \sum_{k=1}^{l+r+1}p^{k-1}\ ^{0}R'^{0}_{q}[\overbrace{^{l}T_{q}^{r}}^{k}(n_t^{i})] \\
&&= \sum_{k=1}^{l+r+1}p^{k-1}\overbrace{^{l}T_{q}^{r}}^{k}[^{0}R'^{0}_{q}(n_t^{i})] = \sum_{k=1}^{l+r+1}p^{k-1}\overbrace{^{l}T_{p}^{r}}^{k}[^{0}R'^{0}_{q}(n_t^{i})]  \nonumber
\end{eqnarray}
where the index $k$ has been relabeled as $k-r-1$, the commutativity of the shift operator, Eq. (\ref{commu}) has been used, by having also Eq. (\ref{reslem5}) in mind, and, finally it has been  used that
\begin{equation}
\overbrace{^{l}T_{q}^{r}}^{k}(x)=\overbrace{^{l}T_{p}^{r}}^{k}(x),
\end{equation}
a consequence of Eq. (\ref{usyful}) when $x$ is an integer $\in [0, p-1]$ [as it is the case with $^{0}R'^{0}_{q}(n_t^{i})$, in consistency with Eq. (\ref{reslem5})]. $\Box$

%Finally, to prove Eq. (\ref{reslem7}) we note that
%\begin{eqnarray}
%&&s_{t+1}^{i} = \sum_{k'=1}^{M}p^{k'-1}n_{t+1}^{i+k'-1}  \\
%&&= \sum_{k'=1}^{M}\sum_{k=1}^{l+r+1}p^{k+k'-2}\overbrace{^{l}T_{q}^{r}}^{k}[^{0}R'^{0}_{q}(n_t^{i+k'-1})]\nonumber \\
%&&= \sum_{k'=1}^{M}\sum_{k=1}^{l+r+1}p^{k+k'-2}\overbrace{^{l}T_{q}^{r}}^{k}[^{0}R'^{0}_{q}(\overbrace{^{l}T_{q}^{r}}^{k'+r}(n_t^{i}))]\nonumber \\
%&&= \sum_{k'=1}^{M}\sum_{k=1}^{l+r+1}p^{k+k'-2}\overbrace{^{l}T_{q}^{r}}^{k}\overbrace{^{l}T_{q}^{r}}^{k'+r}[^{0}R'^{0}_{q}(n_t^{i})]\nonumber \\
%&&= \sum_{k'=1}^{M}\sum_{k=1}^{l+r+1}p^{k+k'-2}\overbrace{^{l}T_{q}^{r}}^{k+k'-1}[^{0}R'^{0}_{q}(n_t^{i})]. \Box\nonumber 
%\end{eqnarray}
%
~\\

\noindent \emph{Corollary: The neighborhood dynamics given by the map Eq. (\ref{NLC}) takes place on a de Bruijn graph.} 
~\\

\noindent \emph{Proof}: From Eq. (\ref{NLC}) we have 
\begin{equation}
\mathbf{d}_{p}(k,n_{t+1}^{i+1})=\overbrace{^{l}T_{p}^{r}}^{k}[^{0}R'^{0}_{q}(n_t^{i+1})] 
\end{equation} 
and also
\begin{eqnarray}
&&\mathbf{d}_{p}(k+1,n_{t+1}^{i})=\overbrace{^{l}T_{p}^{r}}^{k+1}[^{0}R'^{0}_{q}(n_t^{i})] 
=\overbrace{^{l}T_{p}^{r}}^{k}\overbrace{^{l}T_{p}^{r}}^{r+2}[^{0}R'^{0}_{q}(n_t^{i})] \nonumber \\
&&=\overbrace{^{l}T_{p}^{r}}^{k}[^{0}R'^{0}_{q} \overbrace{^{l}T_{q}^{r}}^{r+2}(n_t^{i})] =\overbrace{^{l}T_{p}^{r}}^{k}[^{0}R'^{0}_{q}(n_t^{i+1})]=\mathbf{d}_{p}(k,n_{t+1}^{i+1}) \nonumber \\
\end{eqnarray} 
where commutativity of the shift operator with any CA rule and the recurrence Eq. (\ref{recurri}) have been used.
Thus, we have, from Eq. (\ref{dBm}) $b_{n_{t+1}^{i}n_{t+1}^{i+1}}=1$, which proves the result, since both $t$ and $i$ are arbitrary. $\Box$

There are a number of results that follow from Eq. (\ref{NLC}). We see that the neighborhood dynamics for any rule with given nonvanishing $l$ or $r$, is fully specified by the corresponding $l+r+1$ shift operators acting on a \emph{subset} of all possible local rules $^{0}R^{0}_{q}$. From the latter, only those for which $q=p^{l+r+1}$ and which map the integers $\in [0,q-1]$ to the integers $\in [0, p-1]$ are relevant to describe the neighborhood dynamics of any rule with non-vanishing $l$ or $r$. As an example, out of the $8^{8}=16777216$ local rules $^{0}R^{0}_{8}$, there are only 256 which are relevant to describe the neighborhood dynamics of some  rules with nonvanishing $l$ or $r$. The latter are indeed the 256 $^{1}R^{1}_{2}$ Wolfram rules. 

We also have the following result: local rules $^{0}R^{0}_{q}$ with $q$ a prime number \emph{do not play any role} in describing the neighborhood dynamics of any other rule with non-vanishing $l$ or $r$ (through action of the corresponding shift operators). This follows because when $q$ is prime $q \ne p^{l+r+1}$ for any possible value of $p$, $l$ and $r$, and thus Eq. (\ref{reslem5}) cannot be satisfied.

The spatiotemporal evolution of any CA rule can thus be fully described by means of a de Bruijn graph with $p^{l+r+1}$ vertices and $p^{l+r+2}$ edges by ``coloring'' the vertices with the $p$ possible outputs of the result of the local CA rule $^{0}R^{0}_{q}(n)$ on vertex $n \in [0,p^{l+r+1}-1]$. During the CA evolution, the neighborhood of site $i$ at time $t$ is in the vertex $n_{t}^{i}$. At the same time, the neighborhoods (vertices) at $i-1$, $i$ and $i+1$ are connected through a path of  edges in the forward direction. The colors of the consecutive vertices give the base $p$ representation of the next neighborhood value $n_{t+1}^{i}$: a path connecting $l+r+1$ vertices is thus mapped to one vertex in the graph at the next time step.  

Another result that comes directly from Eq. (\ref{reslem5}) is the following. If a neighborhood value reaches a fixed point (``still life''), i.e., if $n_{t}^{i}=n^{*i}$, then the site $x_{t}^{i}$, to be updated at the next time step in the corresponding local rule, must reach also the constant value $^{0}R'^{0}_{q}(n^{*i})$. This means that a vertex whose color do not match the digit $r+1$ from the left of the base $p$ representation of the vertex cannot be present in a spatial fixed point \cite{Mcintosh}. Therefore a graph giving all possible spatial fixed points (i.e. all possible symbolic strings on the ring that remain constant in time) is obtained by deleting from the de Bruijn graph of the rule those vertices that do not obey the above property.

\begin{figure}
\includegraphics[width=0.525\textwidth]{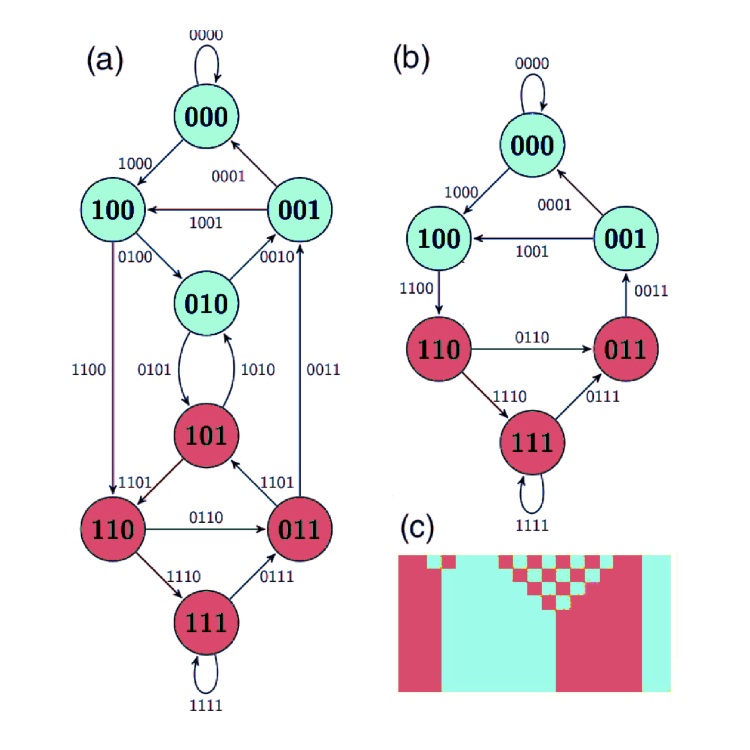}
\caption{(Color online) (a) de Bruijn graph corresponding to rule $^{1}232^{1}_{2}$. (b) Graph giving the fixed points of the spatiotemporal dynamics of rule $^{1}232^{1}_{2}$. (c) Spatiotemporal evolution of rule $^{1}232^{1}_{2}$ in a ring with $N_{s}=20$ and for 10 time steps. Time flows from top to bottom. Light means 0 and dark means 1.}
 \label{dB232}  
\end{figure}

As an example, let us consider Wolfram's rule $^{1}232^{1}_{2}$ with vector $\mathbf{a}=(a_{0},a_{1},...a_{7})=(0,0,0,1,0,1,1,1)$ (From Eq. (\ref{CAW}) this rule has map $x_{t+1}^{i}=x_{t}^{i+1}(x_{t}^{i}+x_{t}^{i-1}-2x_{t}^{i}x_{t}^{i-1})+x_{t}^{i}x_{t}^{i-1}$. The de Bruijn graph corresponding to this rule is shown in Fig. (\ref{dB232})a. The 256 Wolfram rules $p=2$, $l=r=1$ arise from the 256 different vertex colorings of a same ``colorless'' de Bruijn graph, obtained from Eq. (\ref{vertidin}). Vertices colored light correspond to neighborhoods for which the local rule outputs 0, and those colored dark, those for which the local rule outputs 1. Let us consider a path connecting vertices $010 \gets 101 \gets 011 \gets 111$ (i.e. 111 is the starting vertex and 010 the ending vertex). This corresponds to a string '010111' on the ring. Since the colors of the vertices consecutively connected are light $\gets$ dark $\gets$ dark $\gets$ dark, the string '010111' is mapped to a string '0111' at the next time (the previous '1011' block contained in the string '010111' becomes thus updated to '0111').  Since vertices '010' and '101' have a color that does not match the digit at position $r+1=2$ (i.e. the central digit in this case), such vertices cannot belong to a spatial fixed point. In Fig. (\ref{dB232})b these vertices have been eliminated from the previous graph and the resulting graph gives the fixed point structure: all remaining vertices correspond to those strings of zeroes and ones that do not evolve with time. A string of $N_{s}$ edges in this graph is thus a spatial fixed point of the global dynamics. In Fig. (\ref{dB232})c all these observations are confirmed. The spatiotemporal evolution of the rule for an arbitrary initial condition on a ring of $N_{s}=20$ sites is shown. A string of neigborhoods of the form '010' and '101' alternate in time yielding a checkerboard triangular pattern. However, since these neighborhoods do not belong to spatial fixed points, the remaining locations in the ring are driven by the CA rule to a global spatial fixed point that does not contain any such neighborhood '010' o '101', as predicted: The remaining string, which does not longer evolve with time, is a path in the graph of Fig. (\ref{dB232})b. The site values reach this spatial fixed point already with just only four iterations for the initial condition in Fig. (\ref{dB232})c.

\section{Global dynamics of CA}
\label{global}

\subsection{Universal characteristic function}

%\begin{figure}
%\includegraphics[width=0.3\textwidth, angle=270]{Fig3.ps}
%\caption{Scheme clarifying the notation in Sec. \ref{global}: a real number $\phi_{t}\in [0,1]$ is attributed to each iteration of the universal CA map, Eq. (\ref{CA}) by multiplying the site value $x_{t}^{i}$ by a power $p^{i-N_{s}-1}$ and summing over all $i$.  The first two iterations of Wolfram's rule $^{1}170^{1}_{2}$ are then shown for $N_{s}=17$. Starting from the value $\phi_{0}=0.697647094726562$ we have $\phi_{1}=0.395301818847655$, $\phi_{2}=0.790603637695310$, etc. Non-negative integers $I_{t}\equiv p^{N_{s}}\phi_{t}$ which are $\in [0,p^{N_{s}}-1]$ can also be attributed to each iteration step (in the figure, $I_{0}=91442$, $I_{1}=51813$, $I_{2}=103626$, etc.} \label{schem2}  
%\end{figure}

For the local dynamics given by the map in Eq. (\ref{CA}) we also have an associated universal map that governs the $global$ dynamics of the CA. At each time $t$ we can define a (real) number $\phi_{t} \in [0,1)$ that contains all site values as base-$p$ digits as 
%\begin{equation} % for the backward case
%\phi_{t}=\sum_{i=1}^{N_{s}}p^{i-1-N_{s}}x_{t}^{i}
%\end{equation}
\begin{equation}
\phi_{t}=\sum_{i=1}^{N_{s}}p^{i-N_s-1}x_{t}^{i} \label{phi}
\end{equation}
For a given value of all $x_{t}^{i}$'s this real number (in base 10) is unique.  (Alternatively, one can consider a non-negative integer $I_{t} \in [0,p^{N_{s}}-1]$ defined as
%\begin{equation}
%I_{t}=\sum_{i=1}^{N_{s}}p^{i-1}x_{t}^{i} \label{it}
%\end{equation}
$I_{t} \equiv p^{N_{s}}\phi_{t}$.) Note that in the definition, we select site $i=1$ arbitrarily as the "origin" of the order for the powers of $p$. The initial condition at $t=0$ is thus coded as $\phi_{0}$ and the global evolution of the CA is governed by the map
\begin{equation}
\phi_{t+1}=\chi(^{l}R_{p}^{r}; \phi_{t}). \label{returnmap}
\end{equation} 

\begin{figure*}
\includegraphics[width=0.6\textwidth, angle=270]{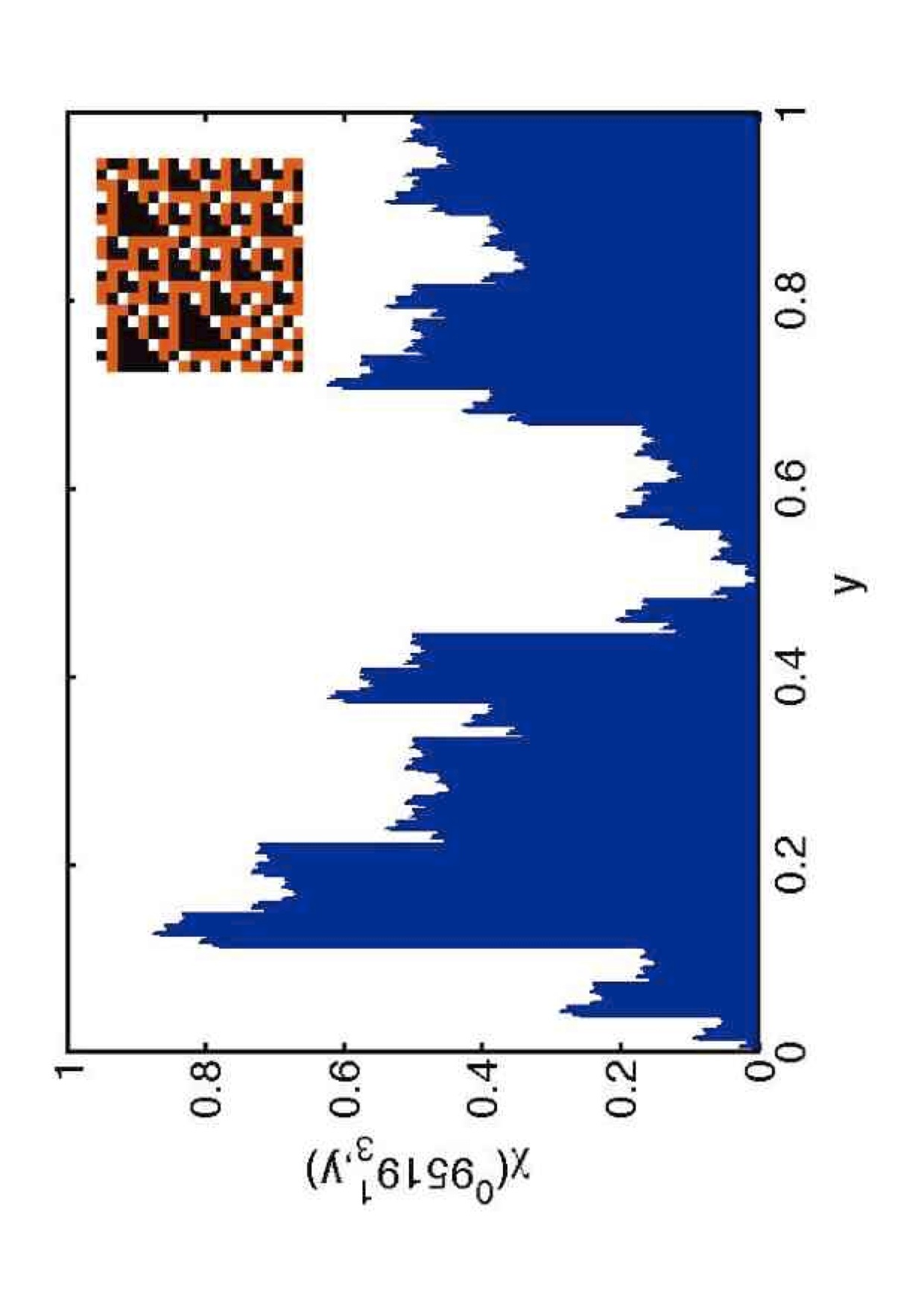}
\caption{(Color online) Characteristic function for CA rule $^{0}9519^{1}_{3}$ calculated from Eq. (\ref{char}) until decimal precision $3^{-6}$ [i.e. $N_{s}=6$ in Eq. (\ref{char})]. The area below the characteristic function is shaded to make clearer the fractality of the curve. The inset shows the spatiotemporal evolution of the rule $^{0}9519^{1}_{3}$ obtained from Eq. (\ref{CA}) for 20 time steps and a ring size $N_{s}=20$ starting from an arbitrary initial condition. Time flows from top to bottom.} \label{rule3char}  
\end{figure*}

\begin{widetext}
Here $\chi(^{l}R_{p}^{r}; y): [0,1] \to [0,1]$ is the characteristic function of CA rule $^{l}R_{p}^{r}$.  From Eqs. (\ref{CA}) and (\ref{phi}) we have:
\begin{eqnarray}
&&\phi_{t+1}= \sum_{i=1}^{N_{s}}p^{i-N_s-1}x_{t+1}^{i} = \sum_{i=1}^{N_{s}}\sum_{n=0}^{p^{r+l+1}-1}p^{i-N_s-1}a_{n}\mathcal{B}\left(n-\sum_{k=-r}^{l}p^{k+r}x_{t}^{i+k}\right) \nonumber \\
&=& \sum_{i=1}^{N_{s}}\sum_{n=0}^{p^{r+l+1}-1}p^{i-N_s-1}a_{n}\mathcal{B}\left(n-\sum_{k=-r}^{l}p^{k+r}\mathbf{d}_p(i+k,p^{N_{s}}\phi_{t})\right) \label{supermap}
% \\
%I_{t+1}&=& \sum_{i=1}^{N_{s}}\sum_{n=0}^{p^{r+l+1}-1}p^{i-1}a_{n}\mathcal{B}\left(n-\sum_{k=-r}^{l}p^{k+r}\mathbf{d}_p(i+k,I_{t})\right) 
\end{eqnarray}
Let $y \in [0,1]$ be a decimal number, the universal characteristic function governing the global dynamics of the map Eq.(\ref{CA}) is given by
\begin{equation}
\boxed{\chi(^{l}R_{p}^{r}; y)=\sum_{i=1}^{N_{s}}\sum_{n=0}^{p^{\rho}-1}p^{i-N_s-1}a_{n}\mathcal{B}\left(n-\sum_{k=-r}^{l}p^{k+r}\mathbf{d}_p(i+k,p^{N_{s}}y)\right)}  \label{char} 
\end{equation}
\end{widetext}

The characteristic function, Eq. (\ref{char}), provides all details of the global evolution of any 1D CA. In Fig. (\ref{rule3char}), it is plotted for the $p=3$, $\rho=2$ CA rule $^{0}9519^{1}_{3}$, calculated from Eq. (\ref{char}) for $N_{s}=6$ (i.e. until decimal precision $3^{-6}$) (in the inset the spatiotemporal evolution of the rule, obtained from Eq. (\ref{CA}) for 20 time steps and a ring size $N_{s}=20$ is shown). The characteristic function contains \emph{fractal structures} \cite{Peitgen}, as can be observed in the figure, and increasing the precision $N_{s}$ merely leads to reproduce the same structures at smaller and smaller scales. The origin of these fractal structures can be understood from the CA evolution, which creates a ``mesoscopic'' scale $\rho=l+r+1$ with an associated nonlocal dynamics (as described in the previous section) which, however, is constrained to satisfy global translation invariance on the ring. 

%In fact, a CA evolving backward in time is similar to a substitution system \cite{VGM4} whose blocks are made to overlap according to a de Bruijn diagram, so that the global evolution is at every time constrained to a sequence of constant length with a total number $N_{s}$ of digits (the proof of this statement will be given elsewhere).  

Global behavior and attractors for all 256 elementary Wolfram's 256 CA rules with 2 symbols and one neighbor to the left and to the right, have been extensively investigated by Wuensche and Lesser \cite{Wuensche} and Chua and his coworkers \cite{Chua1}. Global methods employing characteristic functions have been considered in \cite{Chua2, Chua3} although all these previous works concentrated only in the Wolfram's 256 CA rules. \emph{Eq. (\ref{char}) which defines the universal characteristic function is new and does not contain any adjustable parameter, being also valid for arbitrary alphabet size $p$}.

Interesting information that can be directly drawn from the plot of the characteristic function is the existence of \emph{Gardens of Eden}: i.e. strings that have no preimages in the CA evolution and that, therefore, cannot be reached through the CA dynamics \cite{Moore}. Gardens of Eden can only appear in the spatiotemporal dynamics of a CA rule as initial conditions. A glance at Fig. (\ref{rule3char}) shows us that the interval $0.\overline{8} < \chi(^{0}9519_{3}^{1}; y) < 1$ has no preimage $y$. Since $0.\overline{8}=2\cdot 3^{-1}+2 \cdot 3^{-2}$ this means that, because of the global translation invariance on the ring, all strings containing a block '22' are automatically Gardens of Eden of CA rule $^{0}9519_{3}^{1}$. 

Sometimes it is useful to consider the composition of $\chi(^{l}R_{p}^{r}; y)$ with itself. 
In general, we can define the $\tau$-characteristic function $\chi_{\tau}(^{l}R_{p}^{r}; y)$ as the characteristic function composed with itself $\tau-1$ times, with $\tau$ a natural number. Eq. (\ref{returnmap}) can then be equivalently written as
\begin{equation}
\phi_{t+1}=\chi_{t+1}(^{l}R_{p}^{r}; \phi_{0}) \label{returnmap2}
\end{equation}
i.e. to calculate the global evolution we can either consider the initial state $\phi_{0}$ and its change with time at subsequent time steps, $\phi_{1}$, $\phi_{2}$ etc. by always using the 1-characteristic function Eq. (\ref{char}) or we can fix the global state to $\phi_{0}$ and study the evolution in time of the $t$-characteristic function, acting on $\phi_{0}$. 

%This situation reminds of the Schr\"odinger and Heisenberg pictures of the time evolution of quantum states/operators in quantum mechanics \cite{Sakurai} to which it is analogous. 

\emph{All information} of the global evolution of a CA rule, for a given ring size $N_{s}$, is contained in its \emph{global transition table} $\mathcal{T}\left(^{l}R_{p}^{r}, N_{s}\right)$. First, let us note that Eq. (\ref{returnmap}) can be written alternatively 
\begin{equation}
I_{t+1}=p^{N_{s}}\chi(^{l}R_{p}^{r}; p^{-N_{s}}I_{t}) \label{ret2}
\end{equation} 
in terms of (non-negative) integer numbers $I_{t}=p^{N_{s}}\phi_{t}$. The global transition table of the rule $\mathcal{T}\left(^{l}R_{p}^{r}, N_{s}\right)$ can now be given by using Cauchy's two-line notation (as done with permutation groups). In the upper row, the non-negative integers $I \in [0,p^{N_{s}}-1]$ are listed in increasing order. In the lower row, the non-negative integers $p^{N_{s}}\chi(^{l}R_{p}^{r}; p^{-N_{s}}I)$ (also $\in [0,p^{N_{s}}-1]$) corresponding to the integers $I$ on the first row are listed. We thus have
\begin{widetext}
\begin{equation}
\mathcal{T}\left(^{l}R_{p}^{r}, N_{s}\right)=\begin{pmatrix} 0 & 1 & 2 & ... & p^{N_{s}} -1 \\ \ \ p^{N_{s}}\chi(^{l}R_{p}^{r}; 0) \ \ & \ \ p^{N_{s}}\chi(^{l}R_{p}^{r}; p^{-N_{s}}) \ \ & \ \ p^{N_{s}}\chi(^{l}R_{p}^{r}; 2p^{-N_{s}}) \ \ &\ \ ... \ \ & \ \ p^{N_{s}}\chi(^{l}R_{p}^{r}; 1-p^{-N_{s}})\ \ \end{pmatrix} \label{tran}
\end{equation} 
\end{widetext}
Let us consider as example the rule $^{0}9519_{3}^{1}$ above, on a ring with only $N_{s}=3$ sites. The transition matrix has thus $3^{3}=27$ columns. It can be readily calculated from Eqs. (\ref{char}) and (\ref{tran}) and is equal to
\begin{widetext}
\begin{equation}
\mathcal{T}\left(^{0}9519_{3}^{1}, 3\right)=
\left( \begin{array}{ccccccccccccccccccccccccccc} 0 & 1 & 2 & 3 & 4 & 5 & 6 & 7 & 8 & 9 & 10 & 11 & 12 & 13 & 14 & 15 & 16 & 17 & 18 & 19 & 20 & 21 & 22 & 23 & 24 & 25 & 26 \\  0 & 7 & 4 & 21 & 19 & 19 & 12 & 13 & 13 & 11 & 15 & 13 & 5 & 0 & 1 & 5 & 3 & 4 & 10 & 15 & 13 & 13 & 9 & 10 & 13 & 12 & 13 \end{array} \right)
\end{equation} 
\end{widetext}

A inspection of the table shows that configurations 2, 6, 8, 14, 16, 17, 18, 20, 22, 23, 24, 25, 26 corresponding to the following digit configurations on the ring '002', '020', '022', '112', '121', '122', 
'200', '202', '211', '212', '220', '221', '222' are Gardens of Eden, since they do not appear in the lower row. Some of these Garden-of-Eden configurations (8, 17, 20, 23, 24, 25, 26) had we already detected by inspection of Fig. (\ref{rule3char}) since they all contain the block '22'. We observe that blocks '002' and '112' and their cyclic permutations are also Gardens of Eden. This could also be concluded from Fig. (\ref{rule3char}): by zooming on the corresponding regions we would observe that there is no preimage for these configurations.
We further observe that there are two global attractors for the dynamics: the spatial fixed point 0 ('000') and the 3-cycle formed by configurations $19 \to 15 \to 5$ (i.e. $201 \to 120 \to 012$ on the ring). When the dynamics is confined to motion on this 3-cycle, we see that the corresponding left-shift operator  acting on one of the configurations of the 3-cycle suffices to describe the dynamics, i.e. we simply have
\begin{equation}
x_{t+1}^{i}=\overbrace{^{0}T_{3}^{1}}^{1}(x_{t}^{i})=x_{t}^{i-1} \label{reduced}
\end{equation}
if $n_{t}^{i} \in \{ 19, 15, 5 \}$.  
Since the global transition table has $p^{N_{s}}$ entries, a brute force algorithm to calculate it grows exponentially with system size. The knowledge of the global characteristic function to a certain, not necessarily high, accuracy, can help to drastically accelerate the evaluation of the global transition table, since once all Garden-of-Eden configurations are found, entire parts of the transition table can be evaluated in polynomial $N_{s}$ by simply running the CA starting from any of these configurations until an attractor is reached. The remaining parts of the transition table are then isomorphic to elements of the permutation group. It is to be noted that, since the shift-operator CA rules are bijective when acting on global states of the ring, such rules have no Garden-of-Eden configurations and belong to the permutation group. This observation is intimately related to Cayley's theorem \cite{group}. 

%and to the fact that the site values $x_{t}^{i}$ form a group under the action of any shift operator \cite{group}. In the next subsection we further prove that the shift CA-operators discussed in this article form themselves an abelian group under composition when acting on a ring with $N_{s}=l+r+1$ sites.

\subsection{Global CA}

Let us assume that we are presented with the following problem: we only know the global dynamics through a characteristic function $\chi(^{l}R_{p}^{r}; y)$ and our goal is to find the local CA rule $^{l}R_{p}^{r}$ consistent with the global evolution law. The general problem should lead in many cases to several possible solutions for values of $l$, $r$, $p$ and $R$ which, however, correspond to rules with equivalent dynamical behavior. The more specific question of finding those $R$ with minimal range $l+r+1$ for a given number of symbols $p$ might be intractable in general. Yet, if we restrict ourselves to \emph{global CA}, for which $l+r+1=N_s$, we can show how to construct the local dynamics of the global CA for any $p$ by using the mathematical methods presented in this article. Let us first prove the following result.
~\\

\noindent \emph{Theorem 4: The global shift CA operators (those for which $l+r+1=N_{s}$) form an abelian cyclic group of order $N_{s}$ under composition.}
~\\

\noindent \emph{Proof:} To prove the abelian group structure we have to show that the composition of global shift operators has properties of closure, associativity, existence of identity and inverse elements and commutativity. The latter property follows from Lemma 3, since shift operators are CA rules as well and any CA rule commutes with the shift operators. We consider now the action of global shift operators on the integer number $x^{i} \in [0,p-1]$ with $i \in [1,N_s]$.
~\\
\emph{Closure.} Let $h, k \in [1,N_{s}]$, we have
\begin{eqnarray}
&&\overbrace{^{l}T_{p}^{r}}^{h}[\overbrace{^{l}T_{p}^{r}}^{k}(x^{i})]=\overbrace{^{l}T_{p}^{r}}^{h}(x^{i+k-r-1})  = x^{i+k+h-2r-2} \nonumber \\ && = \overbrace{^{l}T_{p}^{r}}^{m}(x^{i})
\end{eqnarray}
with $m=k+j-r-1$ if $1 \le k+j-r-1 \le N_{s}$ and $m=k+j-r-1-N_{s}$ otherwise (i.e. $m=k+j-r-1 \mod N_s$). This means that the composition of two global shift operators is also a global shift operator with same $l$, $r$ and $p$. Closure  does not hold if $l+r+1 \ge N_{s}$, i.e. if the shift operators are not global.
~\\
\emph{Associative property.} We have
\begin{eqnarray}
&&\overbrace{^{l}T_{p}^{r}}^{h}(\overbrace{^{l}T_{p}^{r}}^{k}[\overbrace{^{l}T_{p}^{r}}^{j}(x^{i})])=
\overbrace{^{l}T_{p}^{r}}^{h}(\overbrace{^{l}T_{p}^{r}}^{k}[x^{i+j-r-1}]) \nonumber \\
&&=\overbrace{^{l}T_{p}^{r}}^{h}(x^{i+j+k-2r-2})=x^{i+j+k+h-3r-3} \nonumber \\
&&=\overbrace{^{l}T_{p}^{r}}^{m}[\overbrace{^{l}T_{p}^{r}}^{j}(x^{i})]=(\overbrace{^{l}T_{p}^{r}}^{h}\overbrace{^{l}T_{p}^{r}}^{k})[\overbrace{^{l}T_{p}^{r}}^{j}(x^{i})]
\end{eqnarray}
where the closure property and the global translation invariance on the ring (i.e. invariance under the transformation $x_t^{i} \to x_t^{i+N_s}$) have been used.
~\\
\emph{Identity element.} From the Corollary accompanying Theorem 2 we have that
\begin{equation}
\overbrace{^{l}T_{p}^{r}}^{r+1}
\end{equation} 
is the identity element.
~\\
\emph{Inverse element.} Since we have
\begin{equation}
\overbrace{^{l}T_{p}^{r}}^{2r+2-m}[\overbrace{^{l}T_{p}^{r}}^{m}(x^{i})]=\overbrace{^{l}T_{p}^{r}}^{m}[\overbrace{^{l}T_{p}^{r}}^{2r+2-m}(x^{i})]=x^{i}=\overbrace{^{l}T_{p}^{r}}^{r+1}(x^{i})
\end{equation}
this means that the operators $\overbrace{^{l}T_{p}^{r}}^{2r+2-m}$ and $\overbrace{^{l}T_{p}^{r}}^{m}$ are the inverse of each other. 

The group is clearly cyclic because there exists at least an element
\begin{equation}
\overbrace{^{l}T_{p}^{r}}^{r}(x^{i}) 
\end{equation}
that generates the whole group. This element has order $l+r+1=N_{s}$ equal to the one of the group.

$\Box$

\noindent \emph{Example:} The Wolfram shift operators: 

\noindent $\overbrace{^{1}T_{2}^{1}}^{1}$ (rule $^{1}170^{1}_{2}$), $\overbrace{^{1}T_{2}^{1}}^{2}$ (rule $^{1}204^{1}_{2}$) and $\overbrace{^{1}T_{2}^{1}}^{3}$ (rule $^{1}240^{1}_{2}$) 
~\\ 
constitute an abelian cyclic group of order $3$ on a ring with $N_s=3$ sites (since then they are global shift operators -i.e global CA rules- as well) but they do not form a group if $N_s \ge 4$. This can be observed from the fact that, for example, the left-shift operator (CA rule $^{1}170^{1}_{2}$) composed with itself, carries a site value $x_t^{i}$ to a site $i+2$ to the left. The resulting action cannot be interpreted as coming from any of the three Wolfram shift operators if $N_s \ge 4$ and thus the closure property is not satisfied.

We exploit the abelian group structure of global shift operators in proving the following result.
~\\

\noindent \emph{Theorem 5: For a global CA rule $^{l}R^{r}_{p}(x_t^{i})$ ($l+r+1=N_{s}$) Eqs. (\ref{NLC}) and (\ref{supermap}) are equal to}
\begin{eqnarray}
\phi_{t+1}&=&\sum_{k=1}^{N_{s}}p^{k-1-N_{s}}\overbrace{^{l}T_{p}^{r}}^{k}[^{0}R'^{0}_{q}\left(p^{N_{s}}\phi_{t}\right)] \label{NLCT6}
\end{eqnarray}

\noindent \emph{Proof:} When $l+r+1=N_{s}$, Eq. (\ref{NLC}) has the form
\begin{eqnarray}
&&n_{t+1}^{i} = \sum_{k=1}^{N_{s}}p^{k-1}\overbrace{^{l}T_{p}^{r}}^{k}[^{0}R'^{0}_{q}(n_t^{i})] 
\label{NLCp}
\end{eqnarray}
Since we have
\begin{eqnarray}
&& n_{t+1}^{i}=\overbrace{^{l}T_{q}^{r}}^{i}(n_{t+1}^{r+1})=
\overbrace{^{l}T_{q}^{r}}^{i}\left(\sum_{k=1}^{N_{s}}p^{k-1}x_{t+1}^{k}\right) \nonumber \\
&&=
\overbrace{^{l}T_{q}^{r}}^{i}\left(p^{N_{s}}\phi_{t+1}\right)
\end{eqnarray}
we obtain, from Eq. (\ref{NLCp})
\begin{eqnarray}
&&\overbrace{^{l}T_{q}^{r}}^{i}\left(p^{N_{s}}\phi_{t+1}\right)=\sum_{k=1}^{N_{s}}p^{k-1}\overbrace{^{l}T_{p}^{r}}^{k}[^{0}R'^{0}_{q}(n_t^{i})] \nonumber \\
&&= \sum_{k=1}^{N_{s}}p^{k-1}\overbrace{^{l}T_{p}^{r}}^{k}[^{0}R'^{0}_{q}\overbrace{^{l}T_{q}^{r}}^{i}\left(p^{N_{s}}\phi_{t}\right)] \nonumber \\
&&= \overbrace{^{l}T_{q}^{r}}^{i}\left(\sum_{k=1}^{N_{s}}p^{k-1}\overbrace{^{l}T_{p}^{r}}^{k}[^{0}R'^{0}_{q}\left(p^{N_{s}}\phi_{t}\right)]\right) 
\end{eqnarray}
where Lemmas 3 and 5 have been used. Now, from Theorem 4, since the inverse element for global shift operators is guaranteed by their group structure, we can operate to both sides of this latter expression with the appropriate inverse global shift operator, i. e.
\begin{eqnarray}
&&\overbrace{^{l}T_{q}^{r}}^{2r+2-i}\overbrace{^{l}T_{q}^{r}}^{i}\left(p^{N_{s}}\phi_{t+1}\right)= \\
&&= \overbrace{^{l}T_{q}^{r}}^{2r+2-i}\overbrace{^{l}T_{q}^{r}}^{i}\left(\sum_{k=1}^{N_{s}}p^{k-1}\overbrace{^{l}T_{p}^{r}}^{k}[^{0}R'^{0}_{q}\left(p^{N_{s}}\phi_{t}\right)]\right) \nonumber
\end{eqnarray}
whence Eq. (\ref{NLCT6}) follows. To prove that Eq. (\ref{supermap}) is also equal to Eq. (\ref{NLCT6}) we observe that
\begin{eqnarray}
&&\phi_{t+1}=\chi(^{l}R_{p}^{r}; \phi_{t}) = \sum_{i=1}^{N_{s}}\sum_{n=0}^{p^{r+l+1}-1}p^{i-N_s-1} \times \nonumber \\
&&\times a_{n}\mathcal{B}\left(n-\sum_{k=-r}^{l}p^{k+r}\mathbf{d}_p(i+k,p^{N_{s}}\phi_{t})\right) \\
&&= \sum_{i=1}^{N_{s}}\sum_{n=0}^{p^{r+l+1}-1}p^{i-N_s-1} a_{n}\mathcal{B}\left(n-\sum_{k=1}^{N_{s}}p^{k-1}\overbrace{^{l}T_{q}^{r}}^{i}(p^{N_{s}}\phi_{t})\right) \nonumber \\
&&= \sum_{i=1}^{N_{s}} p^{i-N_{s}-1}\ ^{0}R'^{0}_{q}[\overbrace{^{l}T_{q}^{r}}^{i}\left(p^{N_{s}}\phi_{t}\right)] \nonumber \\
&&= \sum_{i=1}^{N_{s}} p^{i-N_{s}-1}\overbrace{^{l}T_{p}^{r}}^{i} [^{0}R'^{0}_{q}\left(p^{N_{s}}\phi_{t}\right)]. \qquad \Box \nonumber
\end{eqnarray} 

\noindent \emph{Theorem 6 (Global-local connection): For any continuous real map $\chi: [0,1] \to [0,1]$ of the form} 
\begin{equation}
\varphi_{t+1}=\chi(\varphi_{t}) \label{rmt}
\end{equation} 
\emph{with $\varphi$ a real number $\in [0,1]$, the CA rule given locally by the map}
\begin{eqnarray}
&&x_{t+1}^{i}= \mathbf{d}_{p}\left(i,p^{N_{s}}\chi \left(\sum_{k=1}^{N_{s}}p^{k-N_{s}-1}x_{t}^{k}\right)\right)   \label{map2CA} \\
&&=\left \lfloor\frac{\chi(\sum_{k=1}^{N_{s}}p^{k-N_{s}-1}x_{t}^{k})}{p^{i-N_{s}-1}} \right \rfloor -p \left \lfloor\frac{\chi(\sum_{k=1}^{N_{s}}p^{k-N_{s}-1}x_{t}^{k})}{p^{i-N_{s}}} \right \rfloor \nonumber
\end{eqnarray}
\emph{with $N_{s}=l+r+1$ (and $i \in [1, N_{s}]$) is a diophantine approximation of Eq. (\ref{rmt}) so that, when either $N_{s} \to \infty$ or $p \to \infty$ Eq. (\ref{map2CA}) reproduces exactly Eq. (\ref{rmt}). In the asymptotic limit $p \to \infty$ Eq. (\ref{map2CA}) takes the form}
\begin{eqnarray}
&&x_{t+1}= \left \lfloor p\chi \left(p^{-1}x_{t}\right)  \right \rfloor  \label{asym}
\end{eqnarray}
\emph{with $x_{t} \in [0,p-1] \in \mathbb{Z}$.}
~\\

\noindent \emph{Proof:} By taking into account that
\begin{eqnarray}
&&\varphi_{t+1}=\phi_{t+1}+O(p^{-N_{s}-1}) \approx \phi_{t+1}=\sum_{i=1}^{N_{s}}p^{i-N_{s}-1}x_{t+1}^{i}   \nonumber \\
&&\varphi_{t}=\phi_{t}+O(p^{-N_{s}-1}) \approx \phi_{t}=\sum_{k=1}^{N_{s}}p^{k-N_{s}-1}x_{t}^{k} \label{mapyb}
\end{eqnarray} 
we have, by using Eqs. (\ref{iden}) and (\ref{mapyb}) in Eq. (\ref{rmt}) and equating the coefficients with same powers of $p$ on both sides   
\begin{equation}
x_{t+1}^{i}=\mathbf{d}_{p}\left(i,p^{N_{s}}\phi_{t+1}\right)=\mathbf{d}_{p}\left(i,p^{N_{s}}\chi (\phi_{t})\right) \label{mapy}
\end{equation} 
which proves Eq. (\ref{map2CA}). To prove that this map approximates the original map Eq. (\ref{rmt}) to arbitrary precision we first fix $N_{s}$ finite and take the limit $p \to \infty$. We have, from Eqs. (\ref{limi}) and (\ref{map2CA})
\begin{eqnarray}
 &&\lim_{p \to \infty} \mathbf{d}_{p}\left(i,p^{N_{s}}\chi \left(\sum_{k=1}^{N_{s}}p^{k-N_{s}-1}x_{t}^{k}\right)\right) \\
 && = p^{N_{s}}\chi \left(\sum_{k=1}^{N_{s}}p^{k-N_{s}-1}x_{t}^{k}\right)\ \mathcal{B}\left(i-1\right) \\ 
 && = p^{N_{s}}\chi(\phi_{t})\mathcal{B}\left(i-1\right) \nonumber
\end{eqnarray}
and 
\begin{eqnarray}
 &&\lim_{p \to \infty} \mathbf{d}_{p}\left(i,p^{N_{s}}\phi_{t+1}\right) = p^{N_{s}}\phi_{t+1}\mathcal{B}\left(i-1\right)
\end{eqnarray}
from which we have, since in this limit $\phi_{t+1}=\varphi_{t+1}$ and $\phi_{t}=\varphi_{t}$ (Cauchy convergence of the Diophantine approximation because of the continuity of the map) that $\varphi_{t+1}=\chi(\varphi_{t})$ for $i=1$, thus proving the result. (The same is obtained if one considers finite $p$ and makes the limit $N_{s} \to \infty$ since then every $\varphi$ irrational on the interval can then be reproduced with absolute precision by the CA.) To prove Eq. (\ref{asym}) note that in the limit $p \to \infty$ only $i=1$ is relevant, and hence, we can consider $N_{s}=1$ and drop the unnecessary superindex $i$. Thus, Eq. (\ref{map2CA}) reads in this case
\begin{eqnarray}
&&x_{t+1}=\left \lfloor p\chi(p^{-1}x_{t}) \right \rfloor -p \left \lfloor \chi(p^{-1}x_{t}) \right \rfloor=\left \lfloor p \chi \left(p^{-1}x_{t}\right)\right \rfloor \nonumber
\end{eqnarray}
since $\chi(p^{-1}x_{t})  \in [0,(p-1)/p]$ and, therefore, $\left \lfloor \chi \left(p^{-1}x_{t}\right)\right \rfloor=0$. $\Box$

\subsection{Global CA approach to nonlinear maps on the real line: Application to the logistic map}

The important implication of Theorem 6 is that it provides a direct means to approximate any CA by another (global) one with different $p$, $l$ and $r$ (so that $l+r+1=N_{s}$) to an arbitrary precision, if we know approximately the characteristic function of the former. Furthermore, and what is most important, \emph{it also sistematically allows to find a CA which approximates any real 1D map defined on the interval $[0,1]$ to arbitrary, but fixed, precision.} This leads to interesting insights that we discuss in the following text. 

Let us first recall that Wolfram classified CA behavior phenomenologically into four classes of increasing complexity \cite{Wolfram2}: For a random initial condition a CA evolves into a single homogeneous state (Class 1), a set of separated simple stable or periodic structures (Class 2), a chaotic, aperiodic or nested pattern (Class 3) or complex, localized structures, some times long-lived (Class 4). In a previous work \cite{VGM3} we have shown a simple mechanism to derive the most complex, class 4 CA rules. In this article \emph{we now directly relate Wolfram classes to the qualitative behavior exhibited by real maps}. This is achieved by considering the bifurcation diagram of the corresponding real map. Each parameter regime with different qualitative dynamics can be made to coincide in a one-to-one correspondence with a CA of a certain Wolfram class.

We show now explicitly this correspondence with the logistic map for which one has
\begin{equation}
u_{t+1}=\mu u_{t}(1-u_{t}) \label{logisticmap}
\end{equation}
where $u_{t}$ is a real number defined on the interval $[0,1]$ and $\mu$ is also a real number $\in [0,4]$. From Theorem 6 we have that the following CA
\begin{eqnarray}
&&x_{t+1}^{i}=\left \lfloor\frac{\mu \phi_{t}(1-\phi_{t})}{p^{i-N_{s}-1}} \right \rfloor -p \left \lfloor\frac{\mu \phi_{t}(1-\phi_{t})}{p^{i-N_{s}}} \right \rfloor \label{logisticCA}
\end{eqnarray}
(with $\phi_{t}= \sum_{k=1}^{N_{s}}p^{k-N_{s}-1}x_{t}^{k}$, $i \in [1,N_{s}]$ and $x_{t}^{i} \in [0,p-1]$) integers, approximates the logistic map accurately for $p$ or $N_{s}$ sufficiently large. 

\begin{figure}
\includegraphics[width=0.371 \textwidth, angle=270]{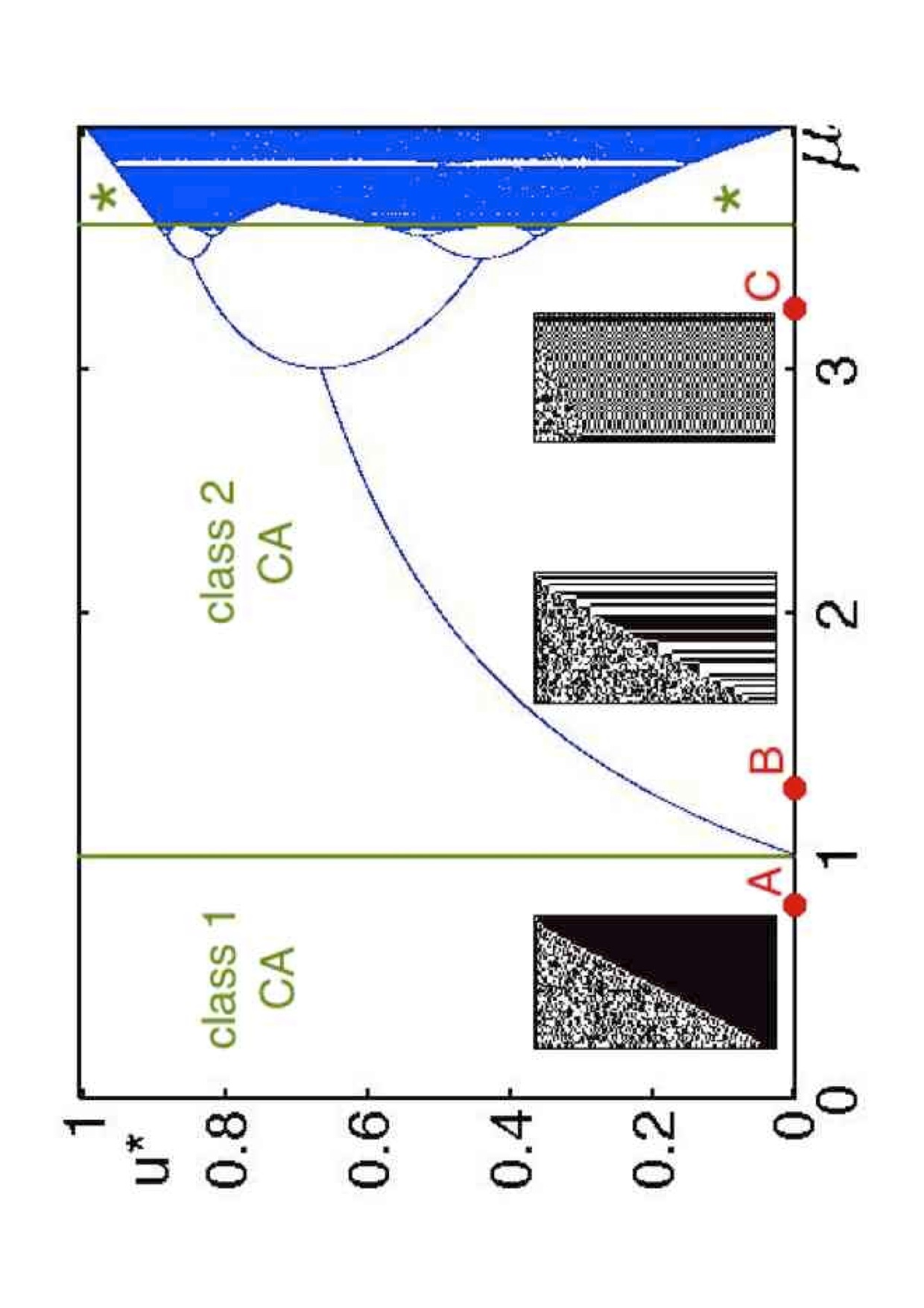}
\caption{(Color online) Bifurcation diagram of the logistic map with the corresponding Wolfram' classes for the logistic CA, Eq. (\ref{logisticCA}) indicated. Inset panels A, B and C:  spatiotemporal evolution of the logistic CA, Eq. (\ref{logisticCA}), for $p=2$ and $N_{s}=50$ and a simple initial condition $x_{0}^{1}=1$ and $x_{0}^{i}=0 \ \forall i >1$ and for values of $\mu$ equal to $0.8$ (A), $1.21$ (B) and $3.2$ (C). Time flows from top to bottom and shown are 150 iteration steps.} \label{bif1}  
\end{figure}

\begin{figure*}
\includegraphics[width=0.7\textwidth, angle=270]{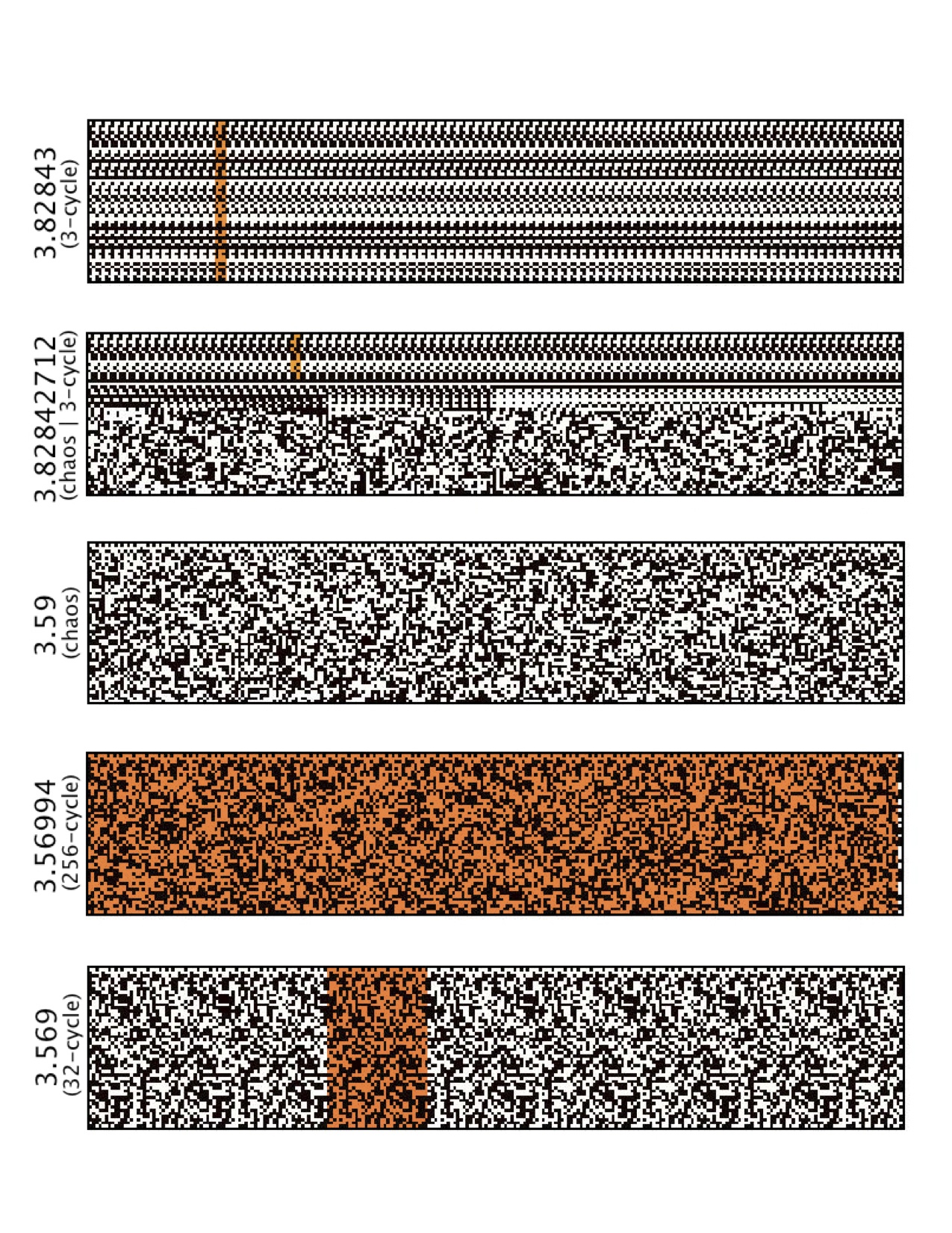}
\caption{(Color online) Spatiotemporal evolution of the logistic CA calculated from Eq. (\ref{logisticCA}) for a time window $t \in [1,257]$ after 110000 iterations on a ring of $N_s=50$ sites, starting from a simple initial condition with 
$x_0^{50}=1$ and $x_0^{i}=0$ for any other value of $i$. Indicated on each panel are the values of $\mu$ used in Eq. (\ref{logisticCA}). Shaded are the periodic structures found over the attractors of the dynamics for each value of $\mu$. The onset of chaos is at $\mu_{\infty}=3.56995$ (not shown).} \label{Pomeau}  
\end{figure*}

In Fig. \ref{bif1} the bifurcation diagram for the logistic map is plotted. In the inset panels, labelled A, B and C, the spatiotemporal evolution of the logistic CA, Eq. (\ref{logisticCA}), is shown, for values of $\mu$ equal to $0.8$, $1.21$ and $3.2$ respectively. We have taken $p=2$ and $N_{s}=50$ and a simple initial condition $x_{0}^{1}=1$ and $x_{0}^{i}=0 \ \forall i >1$. Time flows from top to bottom. For $0 \le \mu < 1$ Eq. (\ref{logisticCA}) exhibits class 1 CA behavior, as shown for the case $\mu=0.8$ in the figure: after a transient, the system evolves into a single, homogeneous state where all site values are zero. For $1 \le \mu < 3.56995$ the behavior is class 2, with Eq. (\ref{logisticCA}) evolving into a set of separated simple stable structures (for $1 \le \mu \le 3.5$, as illustrated by panel B for which $\mu=1.21$ ) or periodic structures (for $3.5 \le \mu \le 3.56995$, as illustrated by panel C, for which $\mu=3.2$). At $\mu=3.56995$  the onset of chaos takes place, and for $3.56995 \le \mu \le 3.82843$ we have the Pomeau-Manneville scenario, with chaotic regimes alternating with windows of multistability. In this region class 3 (chaotic regimes) and class 2 behavior (multistable regimes) alternate. There is, however, a tiny region before the period-3 regime at $\mu=3.82843$ where Class 4 behavior can also be found: within the ring, a coherent structure can be found coexisting with a chaotic region. In Fig. (\ref{Pomeau}) the spatiotemporal evolution of the logistic CA Eq. (\ref{logisticCA}) is shown, in detail for values of $\mu$ corresponding to the period-doubling cascade leading into chaos in the logistic map, the chaotic regime and the multistable, period-3 regime. We observe that the logistic CA is able to capture accurately the dynamics of the (real) logistic map. The logistic CA allows to detect in a glimpse the long periodic behaviors just before the onset of chaos (a 32-cycle and a 256-cycle are shown for the logistic CA, corresponding to the values of $\mu$ in the logistic map where such behaviors are indeed found). Numerical noise is absent in the logistic CA and the precision can be accurately controlled so that long periodic orbits that accurately shadow the real dynamics are rendered accurately in terms of the relevant strings of digits. Of course, because of its finiteness both in alphabet and system sizes the logistic CA has always a trivial $p^{N_{s}}$-cycle and, therefore, only in the limit $N_{s} \to \infty$ or $p\to \infty$ is the logistic map reproduced $exactly$ according to Theorem 6. However, since all realistic computations have finite, limited precision, the global CA obtained from Theorem 6 finely captures the main features deterministic chaos in real maps, even when the dynamics take place on the $N_{s}$ integers $x_{t}^{i} \in [0,p-1]$.

The dynamical behavior found at $\mu=3.82842712$ is very interesting, since it displays the stable coexistence of incoherence and a 3-cycle. This behavior is found within the Pomeau-Manneville scenario just exactly before the stability window where a 3-cycle is observed. Although deeper in the chaotic regime (i.e. $\mu$ slightly lower than 3.82842712) intermittency with chaotic burstings is known to exist, close to the stability regime this $stable$ coexistence of incoherence and regularity takes then place.

% and reminds us of the so-called chimera states \cite{Kuramoto, Abrams, Scholl1, Scholl2, Showalter, gchimeras} where an analogous behavior is observed.  

%It is doubtful whether such behavior has been observed before in the logistic map, because the chaotic part corresponds to the lowest powers of $p$ (the contribution of the chaotic part to the real signal $u_{t} \in [0,1]$ is never larger than $2^{-10}$): The real signal $u_{t}$ plotted versus $t$ appears overall as a period-3 signal because the chaotic behavior is masked in the less significant decimal places. This shows the power of the analysis with the fixed-point arithmetic presented in this paper since fine dynamical details that would otherwise pass unnoticed are revealed through such an approach. 

%Since the CA is here global, the mechanism for these discrete chimeralike states is different to the one discussed in \cite{VGM3}. Moreover, although a nonlocal dynamics of coupled map lattices was shown to display chimera states \cite{Scholl1, Scholl2},  we prove here that they are indeed present in the logistic map itself as well, when one splits the real time series of $u_{t}$ into large symbolic strings from a finite alphabet. This is achieved by the logistic CA, Eq. (\ref{logisticCA}).  Of course, the chimera states found in \cite{Scholl1, Scholl2}, should lead to discrete chimeralike states as well if one uses, instead of the coupled map lattices in \cite{Scholl1, Scholl2}, the corresponding CA derived from Theorem 6 for those coupled map lattices. Further analysis of this statement will be presented elsewhere. 

\section{Conclusions}

We now summarize the main results of this article where the local, nonlocal and global dynamics of CA have been addressed. The results have been derived by means of $\mathcal{B}$-calculus \cite{VGM1, VGM2, VGM3} and of a useful function that can be considered as a ``CA transform'' and which allows to convert an integer number into a string of $N_{s}$ integer digits $\in [0,p-1]$ taken from an alphabet of $p$ symbols. This function is presented in the Lemma 1 of the manuscript and allows to gain insight in CA dynamics. We have also proved a theorem which provides the CA codes of the shift rules in CA space, for every value of $p$ and neighborhood range of the CA rules (Theorem 2). These rules have been shown to be crucial in describing the CA dynamics at the level of entire neighborhoods (i.e. the nonlocal dynamics) and symbolic strings within the ring. The theory has then been related to the use of de Bruijn graphs with colored vertices (Theorem 3). A universal characteristic function for the global dynamics of 1D CA has then been established. As all results derived in this note, the characteristic function does not contain any freely adjustable parameter. Then the group structure of global shift CA operators has been established (Theorem 4). Although shift operators are known to possess group structure in the universe of the integer lattice \cite{group}, the \emph{global CA shift operators} here discussed are a subset of CA rules themselves and only possess group structure on rings of $N_{s}$ sites. One has therefore, besides the group of shift operators acting on the lattice, the group of global CA shift operators acting on themselves. Theorem 5 shows the equivalence of the non-local and global dynamics for global CA. The most important result of this article is Theorem 6, which provides a systematic means to convert any real map to a (global) CA rule. The problem of providing a fully discrete method, a ``theoretical computer arithmetic'', to systematically deal with $any$ real map, as once suggested by McCauley \cite{McCauley}, has thus been tackled here in its wide generality. The advantage of such an approach to deterministic chaos has been made apparent  with the application to the logistic map, since it has lead us to discover interesting dynamical behavior that was subtly hidden within the Pomeau-Manneville scenario, and which displays the spatial coexistence of chaos and regular periodic oscillations.

Support from the Technische Universit\"at M\"unchen - Institute for Advanced Study, funded by the German Excellence Initiative, is gratefully acknowledged.

\end{document}